\documentclass[aps,prb,a4paper,twocolumn,10pt,superscriptaddress,notitlepage]{revtex4-2}
\usepackage[utf8]{inputenc}
\usepackage{graphicx}
\usepackage{amsmath}
\usepackage{natbib}
\usepackage{float}
\usepackage{tikz}
\usepackage{siunitx}
\usepackage{dcolumn}
\usepackage{lmodern}
\usepackage[T1]{fontenc}
\usepackage[unicode=true,
	        colorlinks=true,
	        linkcolor=blue,
	        citecolor=blue,
	        urlcolor=blue]{hyperref} 
\usepackage[cm]{sfmath}
\usepackage{caption}
\DeclareCaptionLabelSeparator{vertline}{ \textnormal{\textbar} }
\usepackage[font=sf, 
            labelfont={sf,bf},
            size=small,
           justification=RaggedRight,
            textfont={sf,sansmath},
            labelsep=vertline
            ]{caption}
\usepackage{ulem}
\usepackage{chemformula}
\usepackage{empheq}
\usepackage{comment}
\usepackage{bm}
\usepackage{gensymb}
\usepackage[top=2.5cm,bottom=2.5cm,left=1.5cm,right=1.5cm]{geometry}
\def\bibsection{\section*{\textsf{References}}} 
\renewcommand{\figurename}{Fig.}
\setcitestyle{super}
\def \pathfigs{}

\DeclareMathOperator\diag{diag}

\newcommand{\papertitle}{\ch{WSe_2} as transparent top gate for near-field experiments}
\begin{document}
\title{\Large\textsf{\papertitle}}
\author{Niels C.H. Hesp}
\affiliation{ICFO-Institut de Ci\`{e}ncies Fot\`{o}niques, The Barcelona Institute of Science and Technology, Av. Carl Friedrich Gauss 3, 08860 Castelldefels (Barcelona), Spain}
\author{Mark Kamper Svendsen}
\affiliation{CAMD, Computational Atomic-Scale Materials Design, Department of Physics, Technical University of Denmark, DK - 2800 Kongens Lyngby, Denmark}
\author{Kenji Watanabe}
\affiliation{Research Center for Functional Materials, National Institute for Materials Science, 1-1 Namiki, Tsukuba 305-0044, Japan}
\author{Takashi Taniguchi}
\affiliation{International Center for Materials Nanoarchitectonics, National Institute for Materials Science,  1-1 Namiki, Tsukuba 305-0044, Japan}
\author{Kristian Sommer Thygesen}
\affiliation{CAMD, Computational Atomic-Scale Materials Design, Department of Physics, Technical University of Denmark, DK - 2800 Kongens Lyngby, Denmark}
\affiliation{Center for Nanostructured Graphene (CNG), Department of Physics, Technical University of Denmark, DK - 2800 Kongens Lyngby, Denmark}
\author{Iacopo Torre}
\email{iacopo.torre@icfo.eu}
\affiliation{ICFO-Institut de Ci\`{e}ncies Fot\`{o}niques, The Barcelona Institute of Science and Technology, Av. Carl Friedrich Gauss 3, 08860 Castelldefels (Barcelona), Spain}
\author{Frank H.L. Koppens}
\email{frank.koppens@icfo.eu}
\affiliation{ICFO-Institut de Ci\`{e}ncies Fot\`{o}niques, The Barcelona Institute of Science and Technology, Av. Carl Friedrich Gauss 3, 08860 Castelldefels (Barcelona), Spain}
\affiliation{ICREA-Instituci\'{o} Catalana de Recerca i Estudis Avan\c{c}ats, 08010 Barcelona, Spain}
\maketitle
\noindent
\textsf{\textbf{
Independent control of carrier density and out-of-plane displacement field is essential for accessing novel phenomena in two-dimensional material heterostructures.
While this is achieved with independent top and bottom metallic gate electrodes in transport experiments, it remains a challenge for near-field optical studies as the top electrode interferes with the optical path.
Here, we systematically characterize the requirements for a material to be used as top-gate electrode, and demonstrate experimentally that few-layer \ch{WSe_2} can be used as a transparent, ambipolar top gate electrode in infrared near-field microscopy.
We perform nano-imaging of plasmons in a bilayer graphene heterostructure and tune the plasmon wavelength using a trilayer \ch{WSe_2} gate, achieving a density modulation amplitude exceeding $2 \cdot 10^{12} ~{\rm cm^{-2}}$.
Moreover, the observed ambipolar gate-voltage response allows to extract the energy gap of \ch{WSe_2} yielding a value of $1.05~{\rm eV}$.
Our results will provide an additional tuning knob to cryogenic near-field experiments on emerging phenomena in 2d-materials and moir\'{e} material heterostructures.
}}
\vspace*{0.2cm}
\noindent
Near-field optical microscopy is a powerful technique for exploring the optical properties of materials on the nanoscale\cite{novotny2006near}. 
Scattering-type scanning near-field microscopy (s-SNOM) is the most commonly used type of near-field optical microscopy in the study of two-dimensional (2D) materials and their heterostructures.
In this configuration a sharp metallic tip (with a typical apex radius of $\approx 20 ~ {\rm nm}$) generates a hotspot of light that interacts with the sample. 
The tip also bridges the momentum mismatch between free-space light and collective excitations, like plasmon \cite{Chen2012,Fei2012} or phonon polaritons\cite{Dai2014, BasovNanophotonics2021}, making s-SNOM an appealing tool for imaging these excitations \cite{Basov2016, Low2017}. 

Importantly, in 2D materials it is possible to continuously tune the carrier density via the field effect, using a gate electrode underneath the sample.
Notably, in single-layer graphene, tuning the carrier density has allowed for understanding of the mechanism of photocurrent generation in near-field photocurrent experiments \cite{Woessner2016}.
Moreover, the carrier density has a direct impact on the plasmon dispersion.
Varying the carrier density in  s-SNOM experiments on graphene plasmons \cite{Chen2012, Fei2012, Woessner2015, Ni2018} has given crucial information to understand the nature of these collective excitations, and the role of many-body effects \cite{Lundeberg2017}. 
Tuning the carrier density also gave an insight into the plasmonic properties of a photonic lattice formed by minimally twisted bilayer graphene \cite{Sunku2018}. 
The modulation of the carrier density can also be exploited to enhance the signal to noise ratio of the recorded signal via a lock-in detection scheme. 
This technique was employed to detect intersubband transitions in transition metal dichalcogenides (TMDs) \cite{Schmidt2018}.
\begin{figure}[t]
\includegraphics{\pathfigs 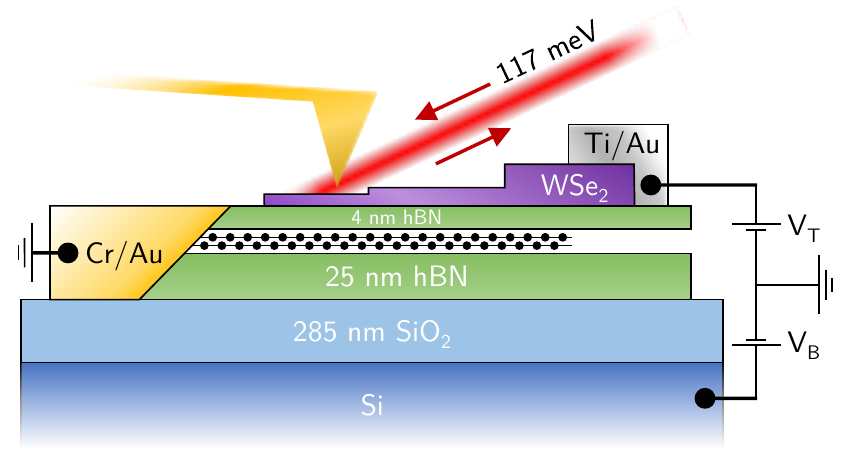}
\caption{\textbf{Device schematic.} Schematic of our near-field experiment performed on a device consisting of hBN-encapsulated bilayer graphene. By applying a voltage to a staircase flake of \ch{WSe_2} (2-6 layers), we tune the carrier density in bilayer graphene without obstructing the near-field access, as verified by probing the plasmon properties. Together with the silicon bottom gate this allows full control of the carrier density and displacement field in bilayer graphene.}
\label{fig:topgate-schematic}  
\end{figure}
The application of a gate voltage on a sample has an additional consequence: beside injecting carriers into the sample, it induces a perpendicular electric displacement field. 
Whilst the effect of the displacement field on the properties of single-layer graphene is minor, it can have a more pronounced impact on multilayer materials. 
For instance, a displacement field leads to a opening of a band gap in bilayer graphene\cite{McCann2013}. 
Hence, independent control of the carrier density and displacement field is of great relevance for exploring novel phenomena in 2D materials using s-SNOM.
For example, domain walls in gapped bilayer graphene are predicted to host long-lived plasmons with lifetimes two orders of magnitude higher than in single-layer graphene \cite{Hasdeo2017}. 
Recent experiments have shown tunability of correlated states in twisted double bilayer graphene using a displacement field \cite{Liu2020, Cao2020}, while twisted trilayer graphene under a displacement field has raised the bar of the superconductivity critical temperature in graphene-based systems beyond $2~{\rm K}$ \cite{Park2021, Hao2021}.  

Independent control of carrier density and displacement field is achieved in transport experiments by placing two separately-contacted gate electrodes, one on the top and one on the bottom of the sample.
This is more difficult in optical experiments, as the top gate electrode interferes with the optical path. 
A sufficiently transparent conducting material is therefore needed for the top electrode. 
Remarkably, the transparency requirement for near-field experiments is even more stringent than for far-field measurements, in particular close to the resonances of the structure, as discussed later.

The traditional choice as a top-gate material for 2D material samples is an evaporated gold film. 
In order to be transparent in the relevant optical range, the electrode thickness must be well below the skin depth of the material to avoid screening of the near-field electromagnetic field. 
Typical metals have a skin depth of tens of nanometres for infrared frequencies \cite{Rahman2016}, which requires the metal film to be only few nanometers thick. Even in this case, as shown in the following, metallic thin films are not able to meet the more stringent requirements imposed by near-field experiments.
On the other hand, conducting oxides, like Indium-Tin Oxide (ITO), are commonly used as transparent electrodes for visible light but are not transparent at infrared frequencies\cite{AMALRICPOPESCU2001139}. 

Recent experiments\cite{Li2020,Sunku2021, Luo2021} have made advances in realizing a transparent top gate for s-SNOM experiments with 2D materials, which also provide atomically flat interfaces and allow for easier device integration. 
Single-layer graphene is sufficiently transparent to probe near-field signals through it \cite{Li2020}, yet its own plasmonic resonance interferes with the optical response of the material underneath, complicating the interpretation of the observed near-field signal \cite{Sunku2021, Luo2021}. 
Still, for studying structural changes that do not involve collective resonances this is not an obstacle \cite{Li2020}. 
As alternative to graphene, the TMD \ch{MoS_2} has been used, which indeed does not host any resonances that disturb the near-field signal \cite{Sunku2021}. 
However, since this material is unipolar due to Fermi level pinning at the contacts, it can only introduce p-type doping in the material below \cite{Kim2017_EF_pinning}.

In this work, we show that \ch{WSe_2} can act as an infrared-transparent bipolar gate electrode for near-field experiments as schematically illustrated in Fig. \ref{fig:topgate-schematic}. 
We validate its performance by studying the change in induced plasmon wavelength $\lambda _{\rm p}$  in bilayer graphene. 
This allows us to determine the carrier density induced by the top gate, while at the same time we characterize to what extent a charged top layer obstructs the observation of plasmons.

We first rigorously derive the transparency requirement for a top-gate electrode to be used in near-field optical experiments. 
The near field response of an homogeneous layered structure is completely determined by its reflection coefficient $r$ for transverse magnetic (TM) waves (the coupling to transverse-electric modes is negligible in near-field) that is a function of the angular frequency $\omega$ and the in-plane wavevector $q$.
Adding another thin layer (either a 2D material or a thin film of a bulk material) on top of the structure alters the reflection coefficient according to (see derivation in Sect. 2 of SI)  
\begin{equation}\label{eq:r'}
r' = r \frac{1 -S -r^{-1}S}{1 +S +rS},
\end{equation}
where $r'$ is the reflection coefficient (function of $\omega$ and $q$) of the structure after the addition of the new layer and $S$ quantifies the optical disturbance introduced by the new layer (as defined in Sect.2 of the SI). 
$S$ depends on the optical conductivity of the new layer as
\begin{equation}\label{eq:S}
S(q,\omega) = \frac{\sigma_{\rm 2D}(q,\omega)\sqrt{\omega^2/c^2-q^2}}{2\epsilon_0\omega},
\end{equation}
where $\sigma_{\rm 2D}$ is the 2D longitudinal (in the direction of the wavevector) conductivity of the material, $c$ is the speed of light, and $\epsilon_0$ the vacuum permittivity.
For 2D materials, $\sigma_{\rm 2D}$ is directly the optical conductivity of the material (in $\Omega^{-1}$), while for thin layers of three-dimensional conducting materials $\sigma_{\rm 2D}= \sigma \delta$, where $\sigma$ is the optical conductivity and $\delta$ is the thickness of the layer (see detailed discussion in Sect.2 of the SI). 
We do not consider thick films since their impact on the optical signal is always stronger than that of thin films made of the same material. 
\onecolumngrid

\begin{figure}
\includegraphics[width =0.3 \textwidth ]{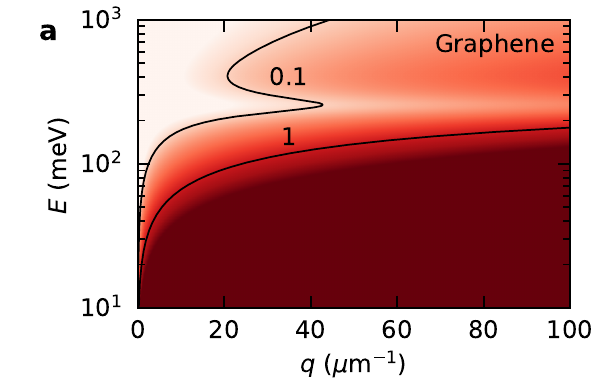}
\includegraphics[width =0.3 \textwidth]{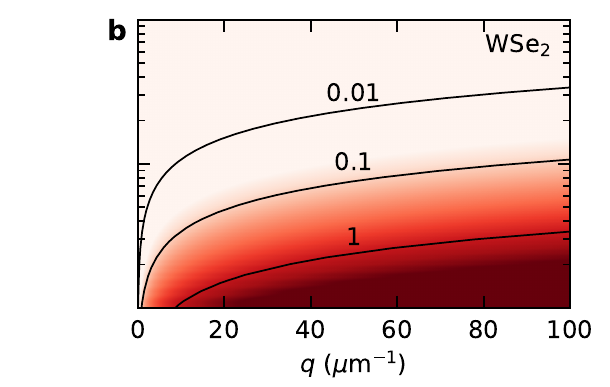}
\includegraphics[width =0.3 \textwidth]{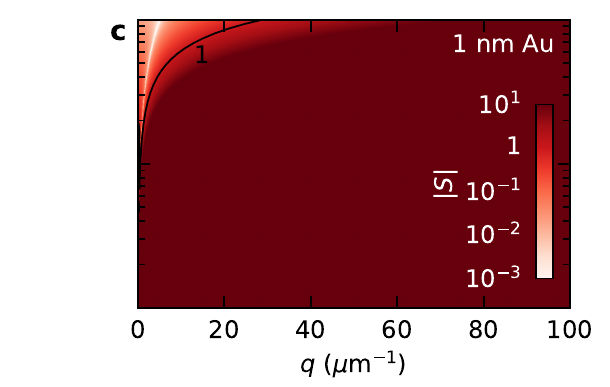}
\caption{\textbf{Optical disturbance introduced by different top layers.} \textbf{a.} Calculated $|S|(q,\omega)$ for single-layer graphene with $n=2 \cdot 10^{12}~ {\rm cm}^{-2}$, $\tau = 200 ~{\rm fs}$ at room temperature. The full non-local longitudinal conductivity has been taken into account. The horizontal feature appearing at $\approx 300 ~{\rm meV}$ is related to the onset of interband transitions at twice the Fermi energy. \textbf{b.} Same as \textbf{a.} for trilayer \ch{WSe_2}, modeled as a Drude conductor with $n=2 \cdot 10^{12}~ {\rm cm}^{-2}$, $m^*= 1.2~ m_{\rm e}$ \cite{Movva2018} $\tau = 100 ~{\rm fs}$ \cite{Movva2015}. \textbf{c.} Same as in \textbf{a.} for a $1 ~{\rm nm}$ thick gold film. Permittivity of gold is taken from Ref. \citenum{Johnson1972}.}
\label{fig:materials_comparison}
\end{figure}
\twocolumngrid
Equation \ref{eq:r'} shows that the reflection coefficient is not affected by the addition of the new layer ($r'\approx r$) when each of the three conditions $|S|\ll 1, |r|,|r|^{-1}$ is satisfied (note that $S = -1$ is the resonance condition of the new layer when isolated, our conditions therefore imply the absence of resonances of the new layer).
The reflection coefficient can be larger than one in absolute value when decaying fields are concerned (i.e. when $q>\omega/c$) and peak exactly at the positions of the collective oscillations\cite{GoncalvesPeres_Book}, reaching values of the order of the quality factor $Q$ of the collective oscillation. 
To observe a collective oscillation without distortion we therefore need $|S|\ll 1/Q$.

Figure \ref{fig:materials_comparison} shows a comparison of the values of $|S|$ calculated, in the ranges of wavevector and frequency that are relevant in typical s-SNOM experiments, for single-layer graphene, trilayer \ch{WSe_2}, and a $1$ nm gold film. 
Most plasmonic excitations probed in room-temperature s-SNOM experiments have $Q\approx 10$. 
It is therefore not possible to study plasmonic excitations using graphene or gold films as a transparent top gate in the mid-infrared regime. In contrast, trilayer \ch{WSe_2} is well suited for this purpose, at least for photon energies above $100$ meV.

The optical disturbance can be reduced by reducing the optical conductivity of the top-gate material, as shown in Eq. \ref{eq:S}. 
For a material with a Drude-like response this can be done by reducing the carrier density and the relaxation time or by increasing the effective mass.
The carrier density is changing when the gate voltage is applied and for effective gating it cannot be smaller than $\approx 10^{12}~{\rm cm^{-2}}$. Additionally, the scattering time $\tau$ is typically too long to play a role for mid-infrared frequencies. 
The search for better materials for top-gating should therefore be oriented towards materials with larger effective masses. 

\ch{WSe_2} can be exfoliated down to a single layer\cite{Podzorov2004, Fang2012} of $0.7~{\rm nm}$ and has relatively low mobilities up to $500~{\rm cm^2/(Vs)}$ \cite{Podzorov2004, Movva2015}. In contrast to \ch{MoS_2}, \ch{WSe_2} is ambipolar and thus allows for injecting both carrier types \cite{Wang2018}. A common issue arising with TMDs is the Schottky barrier forming at the metal-semiconductor interface, typically severely blocking transport through either the valence or conduction band. This can be overcome by a suitable choice of two different metals for the source and drain contact \cite{Das2013}. However, since we intend to use \ch{WSe_2} solely as a gate electrode, a highly resistive contact does not pose an issue, provided we do not modulate the carrier density at high frequencies (Based on an estimated device resistance\cite{Chuang2016} of $\approx 40 ~{\rm k\Omega}$ and a geometrical capacitance of $\approx 0.8~{\rm pF}$ we obtain an RC cut-off frequency of $\approx 200~{\rm MHz}$). 

\begin{figure}[t]
\includegraphics{\pathfigs 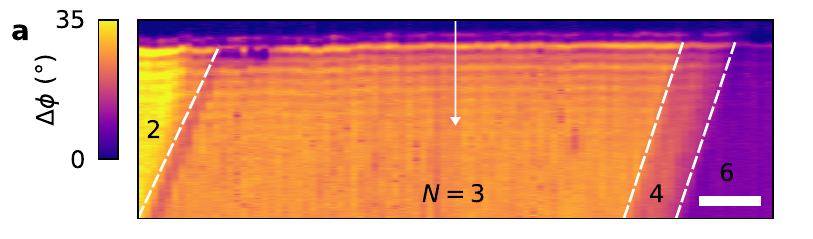}
\includegraphics{\pathfigs 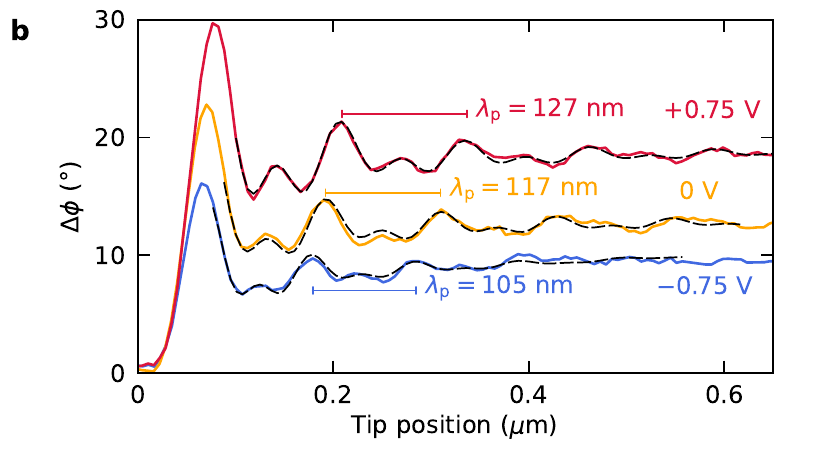}
\caption{\textbf{Controlling the plasmon wavelength with a \ch{WSe_2} top gate.} \textbf{a} Spatial map of the near-field phase contrast at the edge of the bilayer graphene, corresponding to the yellow box in Fig. S1 in Supporting Information. The Silicon bottom gate induces a high carrier density $\sim 10^{13}$~cm$^{-2}$ ($V_{\rm B} - V_{\rm D} = 145~{\rm V}$), allowing the propagation of plasmon polaritons, as seen by the fringes running parallel to the edge. The area shown is covered by \ch{WSe_2} of various thicknesses, as indicated by the number of layers $N$. The excitation energy is $117~{\rm meV}$ and the scale bar is $300~{\rm nm}$. \textbf{b} Line cuts along the white arrow in Fig.~\ref{fig:topgate-spatial}a demonstrate the effect of the \ch{WSe_2} top gate while keeping $V_{\rm B} - V_{\rm D} = 145~{\rm V}$. Without obstructing near-field access, applying a voltage to \ch{WSe_2} alters the carrier density in BLG ($V_{\rm T}$ indicated for each line cut), which affects the measured plasmon wavelength $\lambda _{\rm p}$ as extracted from a fit (black dashed lines).}
\label{fig:topgate-spatial}  
\end{figure}
\onecolumngrid

\begin{figure}[t]
\includegraphics{\pathfigs 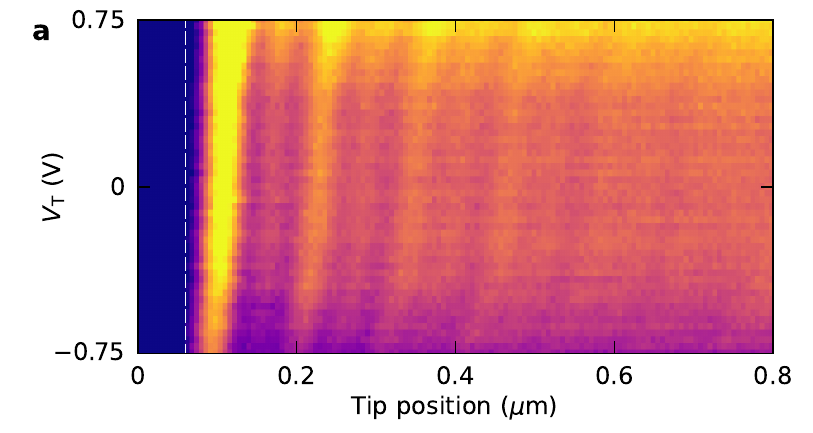}
\includegraphics{\pathfigs 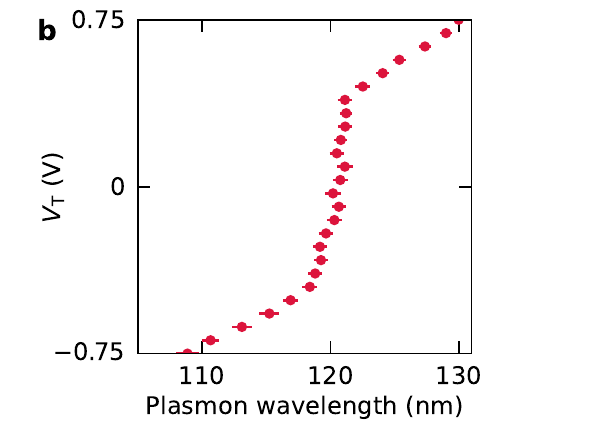}
\includegraphics{\pathfigs 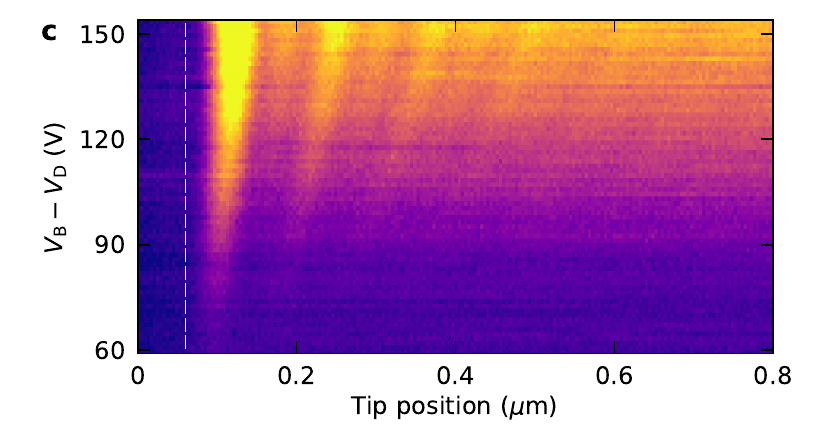}
\includegraphics{\pathfigs 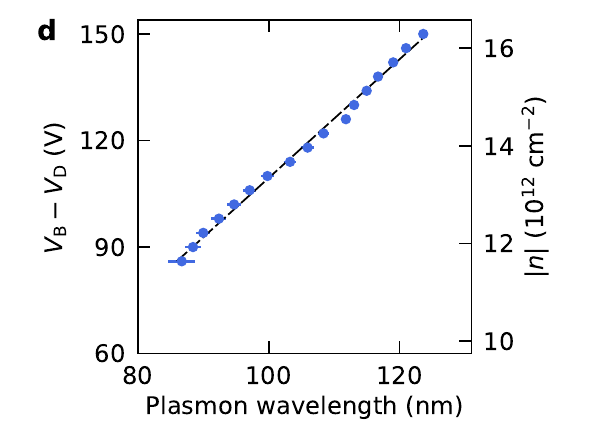}       
\caption{\textbf{Systematically measuring the response of the \ch{WSe_2} top gate.} 
\textbf{a} Line cut of the near-field phase signal along the white arrow in Fig.~\ref{fig:topgate-spatial}a for a range of top gate voltages, while $V_{\rm B} - V_{\rm D} = 145~{\rm V}$. The edge of bilayer graphene is marked with a dashed line. Colour scale is the same as in Fig.~\ref{fig:topgate-spatial}a, covering 18 degree phase difference.
\textbf{b} The extracted plasmon wavelength shows a piece-wise linear dependence on the top gate voltage. The error bars represent $\pm 1 \sigma$. 
\textbf{c} Near-field phase signal along the same line as panel \textbf{a} for a range of bottom gate voltages ($V_{\rm T} = 0~{\rm V}$), serving as calibration to determine the carrier density induced by the \ch{WSe_2} top gate. 
\textbf{d} The extracted plasmon wavelength scales approximately linear with the bottom gate voltage and carrier density, as indicated by a linear fit (dashed line).}
\label{fig:topgate-sweep}
\end{figure}
\twocolumngrid
Figure~\ref{fig:topgate-schematic} shows a schematic of the dual-gated device used in this experiment using \ch{WSe_2} as top gate. 
Our device consists of bilayer graphene (BLG) encapsulated in hexagonal boron nitride (hBN) with a thin \ch{WSe_2} staircase flake acting as a top gate, fabricated as described in the Methods (Sect. 1 of Supporting Information presents an optical image of our device). 
Given the rapid decay of the near-field signal in the out-of-plane direction, we use a $4~{\rm nm}$ thin top hBN flake. 
The \ch{Si/SiO_2} bottom gate serves as a backgate to bring BLG into a highly doped state where plasmons do not suffer from Landau damping \cite{Woessner2015}. 
In addition, the bottom gate provides a reference for determining the plasmon wavelength $\lambda _{\rm p}$ as function of the induced carrier density $n$. Over the course of two months, we did not observe any signs of degradation of the \ch{WSe_2}, despite performing the experiments in ambient conditions\cite{gammelgaard2021}.

Figure~\ref{fig:topgate-spatial}a shows a near-field image of BLG with doping $n \sim 10^{13}~{\rm cm^{-2}}$ induced by the bottom gate at $V_{\rm B}=80~{\rm V}$ along with photodoping, but with $V_{\rm T}=0$. 
Photodoping involves photoexciting defect states at the \ch{SiO_2}/hBN interface \cite{Ju2014, Woessner2016}, which effectively sets the charge-neutrality point $V_{\rm D}$ at $-65~{\rm V}$, as extracted from the maximum of the measured source-drain resistance. The fringes running parallel to the BLG edge are a manifestation of plasmon polaritons, which we observe as both tip-lauched edge-reflected ($\lambda _{\rm p}/2$ period) and edge-launched ($\lambda _{\rm p}$ period) \cite{Woessner2015}, as explained in Methods. 

As a next step, we apply a voltage $V_{\rm T}$ on the \ch{WSe_2} top gate while keeping BLG in the same highly doped state, and record the near-field signal along the arrow in Fig.~\ref{fig:topgate-spatial}a. At this location the top gate consists of three layers of \ch{WSe_2} with total thickness of $\approx 2.2~{\rm nm}$. Figure~\ref{fig:topgate-spatial}b demonstrates that by applying a voltage to the top gate we are able to change the observed plasmon wavelength. By fitting the oscillations to the model introduced in the Methods section, we determine the change in plasmon wavelength to be $\pm 10~{\rm nm}$ for $V_{\rm T} = \pm 0.75~{\rm V}$. This means that \ch{WSe_2} is able to induce carriers of both types in BLG, and thus acts as an ambipolar top gate. In addition, the signal-to-noise ratio of near-field signal stays qualitatively constant while changing the top-gate voltage, confirming that the presence of the top gate does not affect the optical signal.

To study the response of the transparent top gate in more detail, we measure the near-field signal while  systematically scanning the voltage on the top gate, shown in Fig.~\ref{fig:topgate-sweep}a. Judging by the fringe spacing for different $V_{\rm T}$, these data suggest that the top gate only becomes `active' for high $|V_{\rm T}|$. To examine this quantitatively, we fit the data to Eq.~\eqref{eq:snom-fit-function} for each voltage, and extract the plasmon wavelength as function of top gate voltage (Fig.~\ref{fig:topgate-sweep}b). We find indeed a piece-wise linear function with an inactive region for low $|V_{\rm T}|$, while the slope of $\lambda _{\rm p} (V_{\rm T})$ is rather similar for $|V_{\rm T}|>0.4~{\rm V}$. 
\onecolumngrid

\begin{figure}[t]
\centering
\includegraphics{\pathfigs 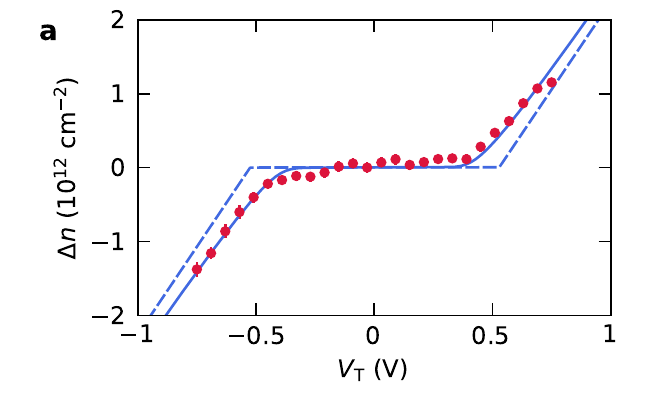}
\includegraphics{\pathfigs 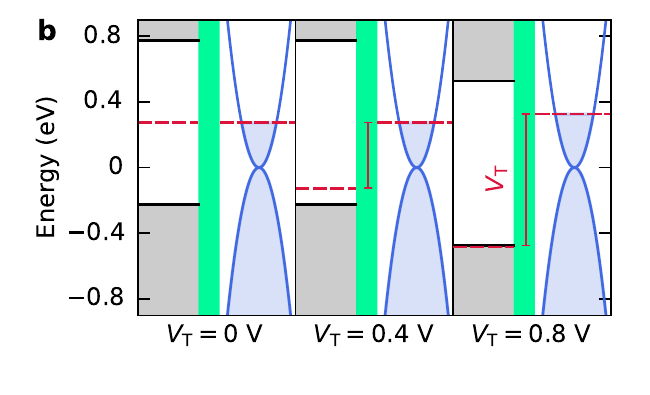}  
\caption{\textbf{Gating efficiency of trilayer \ch{WSe_2} and band alignment.} \textbf{a} Induced shift in the carrier density $\Delta n$ in BLG by applying a voltage to \ch{WSe_2} through a $4~{\rm nm}$ hBN layer. The flat response for small top gate voltages ($|V_{\rm T}| < 0.4~{\rm V}$) arises from the intrinsic gap in \ch{WSe_2}, while for larger voltages ($|V_{\rm T}| > 0.4~{\rm V}$) \ch{WSe_2} acts as a conductor and the slope is determined by the geometric capacitance. From a fit according to the electrostatic model with $T=300~{\rm K}$ we extract a gap of $1.05~{\rm eV}$ in \ch{WSe_2} (solid line). The dashed line represents the calculated $\Delta n$ at zero temperature using the same fit parameters. \textbf{b} Band alignment of semiconducting \ch{WSe_2} (grey bands) with respect to BLG (blue bands) for three different positive top gate voltages. For small $V_{\rm T}$, the chemical potential of \ch{WSe_2} shifts down by $V_{\rm T}$ (middle panel). Once $V_{\rm T}$ is large enough that the chemical potential of \ch{WSe_2} reaches the valence band edge, carriers are injected into BLG moving its chemical potential upwards (right panel). At the same time, the bands of \ch{WSe_2} and BLG are shifted apart such that the $V_{\rm T}$ equals the difference in chemical potentials, indicated by the vertical bar.}
\label{fig:topgate-efficiency}  
\end{figure}
\twocolumngrid
To understand the peculiar shape of $\lambda _{\rm p} (V_{\rm T})$ from a perspective of the electrostatics in our device, we first need to determine the induced carrier density in BLG by the \ch{WSe_2} top gate. To do so, we calibrate $\lambda _{\rm p} (n)$ by using the bottom gate as a reference. According to Eq.~\eqref{eq:electrostatics} introduced below, when $V_{\rm T}=0~{\rm V}$ the induced carrier density can be described by a simple parallel plate capacitor only dependent on $V_{\rm B}$. We measure the plasmon fringes along the same line cut while varying voltage on the bottom gate and keeping $V_{\rm T}=0~{\rm V}$ (Fig.~\ref{fig:topgate-sweep}c,d). 
Applying the same fitting procedure at each voltage point, we find a linear dependence of the plasmon wavelength on $V_{\rm B}$, as is expected for a two-dimensional conductor with parabolic bands \cite{Low2014}. 
Using the capacitance of the bottom gate as mentioned below and $V_{\rm D} = -74~{\rm V}$ in this measurement, we can convert voltages to carrier densities and fit the plasmon wavelength (Fig.~\ref{fig:topgate-sweep}d.) to a linear relation $\lambda _{\rm p} =   a n +b$, yielding $a=8.23\cdot 10^{-12}~{\rm nm\cdot\rm cm^{2}}$ and $b = 34.5~ {\rm nm}$.
From this calibration, we can convert the measured change in plasmon wavelength (Fig.~\ref{fig:topgate-sweep}b) to the carrier density $\Delta n$ induced by the \ch{WSe_2} top gate. 
The final result of our experiment is shown Figure~\ref{fig:topgate-efficiency}a that displays the carrier density induced by the gate as a function of the top-gate voltage.
Note that these measurements were obtained in a completely optical way without relying on transport measurements on graphene.
This demonstrates that graphene plasmons can act as independent local probes of the carrier density that can complement traditional transport measurements.

We can understand our results in terms of a simple model that relates the carrier density $n$ in BLG for given gate voltages $V_{\rm B}$ and $V_{\rm T}$. 
This is based on the equilibrium of the electrochemical potentials (see derivation in Sect. 4 of the Supporting Information) and leads to the relation 
\begin{equation}\label{eq:electrostatics}
n =\frac{C_{\rm B} (V_{\rm B} - V_{\rm D})}{e}+\frac{C_{\rm T}V_{\rm T}}{e} +\frac{C_{\rm T}\Delta\mu_{\ch{WSe_2}}(V_{\rm T})}{e^2} ,
\end{equation}
where $C_{\rm T} \approx 7.7~{\rm mF/m^2}$ and $C_{\rm B} \approx 0.12~{\rm mF/m^2}$ are the geometric capacitances corresponding to the top and bottom gate, $e$ is the unit charge, $\Delta\mu_{\ch{WSe_2}}$ is the shift in chemical potential of \ch{WSe_2} with respect to its value at $V_{\rm T}=0~{\rm V}$, and the quantum capacitance of BLG $C_{\rm Q}\approx 62~{\rm mF/m^{2}}$ has been ignored since it is much larger than $C_{\rm T/B}$. From this equation we see that for $V_{\rm T}=0~{\rm V}$, the carrier density in BLG can be described by the geometric capacitance of the bottom gate (first term), which we used for the calibration of $\lambda _{\rm p} (n)$ above. On the other hand, once we fix $V_{\rm B}$, the change in carrier density is determined by the geometric capacitance of the top gate (second term) and the quantum capacitance of \ch{WSe_2} (last term). The interplay of the last two terms causes the step-like behaviour seen in Fig.~\ref{fig:topgate-efficiency}a, with the central flat region corresponding to the values of $V_{\rm T}$ for which the chemical potential falls deep in the gap of the semiconductor. The calculation of $\Delta\mu_{\ch{WSe_2}} (V_{\rm T})$ is presented in Sect. 4 of the Supporting Information.

A band alignment diagram of \ch{WSe_2} and BLG explains the gating response in more detail (Fig.~\ref{fig:topgate-efficiency}b). Starting with a BLG at a high carrier density induced by the bottom gate, the chemical potentials of \ch{WSe_2} and BLG are aligned for $V_{\rm T}=0~{\rm V}$. In this situation the chemical potential of \ch{WSe_2} resides within the energy gap of \ch{WSe_2}. Upon increasing $V_{\rm T}$, owing to the low quantum capacitance of \ch{WSe_2} in its insulating state, $\Delta\mu_{\ch{WSe_2}}$ shifts down by $V_{\rm T}$ until it reaches the valence band. Once that happens, holes are introduced in the valence band and the capacitively induced electrons in BLG shift the chemical potential of BLG upwards. Due to the relatively high density of states of the valence band of \ch{WSe_2}, its chemical potential remains close to the valence band edge for higher carrier densities. Since the difference in chemical potential between \ch{WSe_2} and BLG has to be equal to $V_{\rm T}$, the bands of \ch{WSe_2} and BLG separate in energy for higher carrier densities.

In fitting the data to our model in Fig.~\ref{fig:topgate-efficiency}a, we only leave the valence and conduction band energies as free fit parameters, while the other parameters can be estimated with sufficient accuracy. The fit yields a band gap of $(1.05\pm 0.02)~{\rm eV}$ (the error being estimated from the fitting procedure), and is almost perfectly centred at $V_{\rm T}=0~{\rm V}$. 
We theoretically calculated the electronic band gap and the densities of states at the band extrema of our material by performing Density Functional Theory (DFT) calculations for trilayer \ch{WSe_2} using the Atomic Simulations Recipe environment and the GPAW package\cite{gjerding2021atomic,enkovaara2010electronic}. The electronic band structure and the densities of the states are reported in the S.I. 
We find an electronic band gap of $1.05 ~{\rm eV}$ which agrees very well with the experimental results in this work and previous theoretical calculations\cite{Movva2018,Dai2015_bandstr,Javaid2018}. 
The agreement with the experiment is surprising because DFT is known to underestimate the band gaps, and generally the experimental band gap of trilayer \ch{WSe_2} is reported \cite{Zhao2013, Zeng2013, Movva2018} at higher values around $1.45~{\rm eV}$. We believe that this discrepancy is caused by the charge imbalance present in \ch{WSe_2} when generating an external electric field. 
The internal field due to the charge imbalance can modify the electronic bands and is expected to reduce the gap size due to band bending.

In summary, we have shown that few-layer \ch{WSe_2} can be used as a transparent ambipolar top gate for near-field experiments. 
This is demonstrated by tuning the plasmonic excitations in bilayer graphene via a \ch{WSe_2} top gate without obscuring the near-field scattering signal. 
Nanoscale measurements of the plasmon wavelength allow us to extract the gating efficiency, which we capture in a minimal model that considers the geometric and quantum capacitances. 
We expect other members of the TMD class, in particular those with higher effective masses, to be equally suitable as infrared transparent top gates due to their similarity, while their scalability via CVD growth allows for easy device integration \cite{Lin2014, Wang2014, Eichfeld2015, Liu2015}. 
This work paves the way for future cryogenic near-field experiments on exotic states in dual-gated sample geometries \cite{Hasdeo2017, Liu2020, Cao2020, Park2021, Hao2021}.
On the other hand, our experiment allowed to directly probe the energy gap and band-alignment (relative to graphene) of semiconducting 2D materials, providing information that can complement the data obtained by other techniques \cite{morpurgo2021}.

\section*{Methods}
\subsection*{Device fabrication}
To fabricate the hBN~($25~{\rm nm}$)/BLG/hBN~(4~nm)/\ch{WSe_2} heterostructure depicted in Fig.~\ref{fig:topgate-schematic}, we used a standard stacking method \cite{Zomer2014}. 
First, we exfoliate from \ch{WSe_2} crystals (HQ Graphene) a thin staircase flake that acts as a top gate. 
Making use of a stamp of polydimethylsiloxane (PDMS) covered by a thin polycabonate film, we pick up the \ch{WSe_2} flake, followed by the hBN and BLG flakes. These steps are performed at $40~^\circ{\rm C}$, while the final stack is deposited on the \ch{Si/SiO_2} target substrate at $165~^\circ{\rm C}$ to assist in squeezing out any bubbles \cite{Purdie2018}. 
Low-resistance contacts to BLG are made by reactive ion etching in a \ch{CHF_3}/\ch{O_2} gas mixture, followed by \ch{Cr/Au} metallization \cite{Wang2013}. The \ch{WSe_2} flake is contacted with Ti/Au with the aim to avoid a high Schottky barrier. 
However, we found in other devices that \ch{Cr/Au} provides an equally well-functioning contact to a \ch{WSe_2} top gate. 
Finally, to prepare the device for s-SNOM measurements, we mechanically cleaned the top surface using contact-mode Atomic Force Microscope (AFM) brooming \cite{Goossens2012}.
\subsection*{Measurement details}
The near-field measurements have been performed on the neaSNOM platform (neaspec), equipped with a \ch{CO_2} gas laser (Access Laser) and a fast \ch{HgCdTe} detector (Kolmar Technologies). 
We focus $15~{\rm mW}$ of infrared light (wavelength $10.6~{\rm \mu m}$, corresponding to a photon energy of $117~{\rm meV}$) onto a \ch{PtIr}-coated AFM tip (Nanoworld), which oscillates at a frequency $\approx 250~{\rm kHz}$ with a tapping amplitude of $80-100~{\rm nm}$. 
We operate the system in a pseudoheterodyne mode \cite{Ocelic2006} using a \ch{ZnSe} beam splitter to obtain the phase resolved near-field signal. 
To avoid detecting unwanted far-field signals, we record the near-field signal at the third harmonic of the cantilever oscillation. All measurements are performed in ambient conditions.
\subsection*{Extraction of plasmon wavelength}
The hotspot at the tip interacts with the charge carriers in graphene and produces collective excitations that are reflected by interfaces, return to the tip, and are finally converted into a scattered field, as measured by the infrared photodetector. This leads to oscillations in the near-field signal, as seen in Figure~\ref{fig:topgate-spatial}. 
Because plasmons make a round trip, peaks in the near-field signal occur at half the plasmon wavelength, satisfying constructive interference underneath the tip. 
In addition to these tip-launched plasmons, the graphene edge can also act as a launcher. 
In that case, the near-field signal underneath the tip is modulated at a spacing equal to the plasmon wavelength. 
We determine the plasmon wavelength $\lambda _{\rm p} = 2\pi/q_1$ using the model introduced in Ref.~\citenum{Woessner2015}, which accounts for both contributions in the optical signal:
\begin{equation}\label{eq:snom-fit-function} 
s_{\rm opt}(x) = A \frac{e^{i2qx}}{\sqrt{x}} + B \frac{e^{iqx}}{x^\alpha} + Cx + D ,
\end{equation}
with $x$ as the distance from the graphene edge to the tip position, and $q=q_1+iq_2$ as the complex plasmon wavevector. 
The first term describes tip-launched plasmons that decay on a scale $\propto 1/q_2$, and takes into account a geometrical decay factor $\sqrt{x}$. 
The second term accounts for edge-launched plasmons with a variable decay factor $\alpha \sim 1$. $A$, $B$ are complex fit parameters, and $Cx + D$ captures any offsets in the signal.


\section*{Acknowledgements}
F.H.L.K. acknowledges financial support from the Government of Catalonia trough the SGR grant, and from the Spanish Ministry of Economy and Competitiveness, through the Severo Ochoa Programme for Centres of Excellence in R\&D (Ref. SEV-2015-0522), and Explora Ciencia  (Ref. FIS2017-91599-EXP). 
F.H.L.K. also acknowledges support by Fundacio Cellex Barcelona, Generalitat de Catalunya through the CERCA program, and the Mineco grant Plan Nacional (Ref. FIS2016-81044-P) and the Agency for Management of University and Research Grants (AGAUR) (Ref. 2017-SGR-1656).  Furthermore, the research leading to these results has received funding from the European Union's Horizon 2020 programme under grant agreements Refs. 785219 (Graphene Flagship Core2) and 881603 (Graphene Flagship Core3), and Ref. 820378 (Quantum Flagship). 
This work was supported by the ERC TOPONANOP under grant agreement Ref. 726001.
N.C.H.H. acknowledges funding from the European Union's Horizon 2020 research and innovation programme under the Marie Skłodowska-Curie grant agreement Ref. 665884.
I.T. acknowledges funding from the Spanish Ministry of Science, Innovation and Universities (MCIU) and State Research Agency (AEI) via the Juan de la Cierva fellowship Ref. FJC2018-037098-I.
K.W. and T.T. acknowledge support from JSPS KAKENHI (Grant Numbers 19H05790, 20H00354 and 21H05233).
K.S.T. acknowledge funding from the European Research Council (ERC) Grant No. 773122 (LIMA). 
K.S.T. is a Villum Investigator supported by VILLUM FONDEN (grant no. 37789).

\section*{Author contributions}
N.C.H.H. and F.H.L.K. conceived the experiment.
N.C.H.H. fabricated the devices and performed the experiments. 
K.W. and T.T. synthesized the hBN crystals.
I.T. developed the theoretical modeling of the experiment.
M.K.S. and K.S.T. performed the DFT band-structure calculations.
N.C.H.H., I.T. and F.H.L.K. analysed the results and wrote the manuscript with input from all the authors.
F.H.L.K. supervised the work.

\section*{Competing Financial Interests}
The authors declare no competing financial interests.

\section*{Data Availability Statement}
The data that support the plots within this paper and other findings of this study are available from the corresponding author upon reasonable request.

\def\bibsection{\section*{\textsf{Supplementary references}}} 

\renewcommand{\figurename}{Fig.}
\setcounter{equation}{0}
\setcounter{figure}{0}
\setcounter{table}{0}
\makeatletter
\renewcommand{\theequation}{S\arabic{equation}}
\renewcommand{\thefigure}{S\arabic{figure}}
\renewcommand{\thetable}{\arabic{table}}
\renewcommand{\bibnumfmt}[1]{[S#1]}
\renewcommand{\citenumfont}[1]{S#1}

\onecolumngrid
\clearpage 
\noindent
\textbf{\Large \textsf {Supporting information for ``\papertitle"}}
\vspace{\columnsep}

\section{Device geometry}
\label{sec:app-geometry}
\begin{figure}[h]
\centering
\includegraphics{\pathfigs  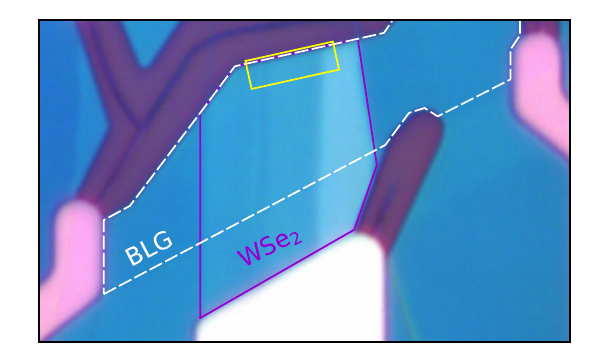}
\caption{\textbf{Optical image of our device.} The locations of bilayer graphene and \ch{WSe_2} flakes are indicated by a dashed white line and a solid purple line respectively.}
\label{fig:topgate-optical}  
\end{figure}

Figure~\ref{fig:topgate-optical} shows an optical image of the device featured in the main text. 
The stack was etched in the dark purple regions, providing a reflecting interface for plasmons. 
The yellow rectangle corresponds to the scan of Fig.~3a in the main text.
\section{Reflection perturbation induced by an additional layer}
We consider the transmission of electromagnetic fields inside homogeneous layered structures made of uniaxial dielectrics with the special axis oriented in the stacking direction $\hat{\bm z}$.
The description of electromagnetic fields can be separated in Transverse Electric (TE) and Transverse Magnetic (TM) modes, with only the latter having a significant coupling to near-field probes.
Defining the direction of in-plane propagation as $\hat{\bm x}$, TM fields in homogeneous planar structures can be conveniently described in terms of the vector 
\begin{equation}
\bm v(z) = 
\begin{pmatrix}
E_x(z) \\ Z_0 H_y(z)
\end{pmatrix},
\end{equation}
where $\bm E$ is the electric field, $\bm H$ is the magnetic field, $Z_0 =\sqrt{\mu_0/\epsilon_0}\approx 377~{\rm \Omega}$ is the vacuum impedance, and all the fields have the form $\bm F(\bm r, t) = \bm F(z)e^{iqx-i\omega t}$.
With this notation Maxwell's equations reduce, in a uniform layer and for the relevant polarization, to the first-order differential equation 
\begin{equation}\label{eq:maxwell}
\frac{d}{dz} \bm v(z) = iK \bm v(z),
\end{equation}
where
\begin{equation}
K =  
\begin{pmatrix}
  0 & \frac{cq_z^2}{\omega\epsilon_\parallel} \\
 \frac{\omega}{c}\epsilon_{\parallel} & 0 \\
\end{pmatrix}
\end{equation}
and
\begin{equation}
q_z =\xi \sqrt{\frac{\omega^2}{c^2}\epsilon_\parallel \mu_\parallel-\frac{\epsilon_\parallel}{\epsilon_\perp}q^2 }.
\end{equation}
Here the optical constant of the material are in the form ${\bm \epsilon} = \diag (\epsilon_\parallel, \epsilon_\parallel, \epsilon_\perp)$, and ${\bm \mu} = \diag (\mu_\parallel, \mu_\parallel, \mu_\perp)$, $q$ is the in-plane wave vector, and the sign $\xi=\pm 1$ is chosen such that $\Im m [q_z]\geq 0$ and, in case $\Im m [q_z]= 0$, $\Re e [q_z]\geq 0$.
The solution of (\ref{eq:maxwell}) in a uniform layer allows to relate the fields on the top of the layer to the ones at the bottom of the layer as
\begin{equation}
\bm v_{\rm top}
=
M_{\rm layer} \cdot 
\bm v_{\rm bottom},
\end{equation}
where $M_{\rm layer}$ is the transfer matrix of the layer, which reads
\begin{equation}\label{eq:layer}
M_{\rm layer} = \exp[i\delta K] = \begin{pmatrix}
\cos(q_z\delta) & i \sin(q_z\delta)\frac{cq_z}{\omega\epsilon_\parallel}\\
i \sin(q_z\delta)\frac{\omega\epsilon_\parallel}{cq_z} & \cos(q_z\delta)
\end{pmatrix}
\end{equation}
is the transfer matrix of the layer, with $\delta$ being the layer thickness.

A homogeneous layer supports two independent solutions corresponding to the two eigenvectors of $K$ with corresponding eigenvalues $\pm q_z$, corresponding to the upwards (downwards) propagating wave. 
The general solution reads
\begin{equation}\label{eq:generalsolution}
\bm v (z) = A_+ e^{iq_z z}
\begin{pmatrix}
1\\
\frac{\omega\epsilon_{\parallel}}{cq_z}
\end{pmatrix}
+A_- e^{-iq_z z}
\begin{pmatrix}
1\\
-\frac{\omega\epsilon_{\parallel}}{cq_z}
\end{pmatrix}
\end{equation}
with $A_\pm$ being the amplitudes of the two waves.

In a similar way the fields on the top and on the bottom of a two-dimensional conducting, non-magnetic sheet are connected by the boundary conditions on the electric and magnetic field.
This relation can be expressed as
\begin{equation}
\bm v_{\rm top}
=
M_{\rm sheet} \cdot 
\bm v_{\rm bottom},
\end{equation}
with $M_{\rm sheet}$ being the transfer matrix of a 2D sheet
\begin{equation}\label{eq:sheet}
M_{\rm sheet} = 
\begin{pmatrix}
1 &  0 \\
- \sigma^{\rm 2D}Z_0  & 1 
\end{pmatrix}.
\end{equation}
The total transfer matrix of a structure composed of many layers and sheets is given by the matrix product of the individual layers $M_{\rm structure} = M_N \cdot M_{N-1} \dots\cdot M_1$.

The reflection $r$ and transmission $t$ coefficients (for the electric field) can be obtained by imposing scattering boundaries conditions.
This means that above the structure the field is given by an incident wave propagating towards $-z$ and a reflected wave propagating towards $+z$, while underneath the structure there is only a transmitted wave propagating towards $-z$. Making use of (\ref{eq:generalsolution}) and relating the field on the top and bottom of the structure via the total transfer matrix we obtain
\begin{equation}\label{eq:r}
\begin{pmatrix}
1\\
-\frac{\omega\epsilon_{\parallel}^{\rm T}}{cq_z^{\rm T}}
\end{pmatrix}
+
r 
\begin{pmatrix}
1\\
\frac{\omega\epsilon_{\parallel}^{\rm T}}{cq_z^{\rm T}}
\end{pmatrix}
 = 
t  M_{\rm structure} \cdot \begin{pmatrix}
1\\
-\frac{\omega\epsilon_{\parallel}^{\rm B}}{cq_z^{\rm B}}
\end{pmatrix}
=t\begin{pmatrix}
a\\b
\end{pmatrix},
\end{equation}
where the superscripts T and B denote the medium that is on the top and on the bottom of the structure.

In the same way after adding a new element to the structure characterized by a transfer matrix $M'$ we can obtain the modified reflection and transmission coefficients $r'$ and $t'$ from
\begin{equation}\label{eq:r1}
\begin{pmatrix}
1\\
-\frac{\omega\epsilon_{\parallel}^{\rm T}}{cq_z^{\rm T}}
\end{pmatrix}
+
r '
\begin{pmatrix}
1\\
\frac{\omega\epsilon_{\parallel}^{\rm T}}{cq_z^{\rm T}}
\end{pmatrix}
 = 
t' M'\cdot M_{\rm structure} \cdot  \begin{pmatrix}
1\\
-\frac{\omega\epsilon_{\parallel}^{\rm B}}{cq_z^{\rm B}}
\end{pmatrix}
=t'\begin{pmatrix}
a'\\b'
\end{pmatrix}.
\end{equation}
We can then relate $r'$ to $r$ by comparing (\ref{eq:r}-\ref{eq:r1}) and using
\begin{equation}
\frac{a'}{b'} = \frac{M_{11}'\frac{a}{b}+M_{12}'}{M_{22}'+M_{21}' \frac{a}{b}},
\end{equation}
from which we obtain
\begin{equation}\label{eq:r1r}
\frac{r'}{r} = \frac{\frac{1}{2}\left(M_{11}'+M_{22}'+\frac{\omega \epsilon_\parallel^{\rm T}}{cq_z^{\rm T}}M_{12}' + \frac{cq_z^{\rm T}}{\omega \epsilon_\parallel^{\rm T}} M_{21}'\right)
+r^{-1}\frac{1}{2}\left(M_{11}'-M_{22}'-\frac{\omega \epsilon_\parallel^{\rm T}}{cq_z^{\rm T}}M_{12}' + \frac{cq_z^{\rm T}}{\omega \epsilon_\parallel^{\rm T}} M_{21}'\right)}
{\frac{1}{2}\left(M_{11}'+ M_{22}'-\frac{\omega \epsilon_\parallel^{\rm T}}{cq_z^{\rm T}}M_{12}' - \frac{cq_z^{\rm T}}{\omega \epsilon_\parallel^{\rm T}} M_{21}'\right)
+r\frac{1}{2}\left(M_{11}'- M_{22}'+\frac{\omega \epsilon_\parallel^{\rm T}}{cq_z^{\rm T}}M_{12}' - \frac{cq_z^{\rm T}}{\omega \epsilon_\parallel^{\rm T}} M_{21}'\right)}.
\end{equation}
This equation holds in general and relates the reflection coefficient in the presence of the additional layer $r'$ to the original reflection coefficient of the structure $r$.
Specializing (\ref{eq:r1r}) to a two-dimensional sheet with $M'$ given by (\ref{eq:sheet}) yields Eq. (1) of the main text
\begin{equation}\label{eq:2d}
\frac{r'}{r} = \frac{1-S -Sr^{-1}}
{1 + S +rS},
\end{equation}
with
\begin{equation}\label{eq:S3d}
S =\frac{cZ_0 \sigma^{\rm 2D}q_z^{\rm T}}{2\omega \epsilon_\parallel^{\rm T}},
\end{equation}
that reduces to Eq. (2) of the main text if the material on top of the structure is vacuum.

For a 3D layer with a finite thickness $\delta$ the transfer matrix $M'$ is given by (\ref{eq:layer}) and the general formula (\ref{eq:r1r}) reduces to 
\begin{equation}\label{eq:3d}
\frac{r'}{r} = \frac{1-S\left[1+\left(\frac{q_z\epsilon_\parallel^{\rm T}}{q_z^{\rm T}\epsilon_\parallel}\right)^2\right] -Sr^{-1} \left[1-\left(\frac{q_z\epsilon_\parallel^{\rm T}}{q_z^{\rm T}\epsilon_\parallel}\right)^2\right]}
{1 + S\left[1+\left(\frac{q_z\epsilon_\parallel^{\rm T}}{q_z^{\rm T}\epsilon_\parallel}\right)^2\right] +rS\left[1-\left(\frac{q_z\epsilon_\parallel^{\rm T}}{q_z^{\rm T}\epsilon_\parallel}\right)^2\right]},
\end{equation}
with $S$ given by
\begin{equation}\label{eq:S3D}
S = -\frac{i}{2}\tan(q_z\delta)\frac{q_z^{\rm T}\epsilon_\parallel}{q_z\epsilon_\parallel^{\rm T}}.
\end{equation}
When near-field is concerned ($q\gg\omega/c$) and for an isotropic, non-magnetic material, with optical conductivity $\sigma$ that is sufficiently large ($|\sigma| \gg \omega \epsilon_0$, as it is the case for metals below their plasma frequency) the quantities in square brackets in (\ref{eq:3d}) are close to one, recovering the same form of (\ref{eq:2d}). 
If moreover the thin-film condition $|q_z|\delta \ll 1$ is fulfilled (and again $|\sigma|\gg\omega \epsilon_0$) Eq. (\ref{eq:S3D}) also reduces to Eq. (\ref{eq:S3d}) with a 2D conductivity given by $\sigma^{\rm 2D} =\delta \sigma$.
\section{Energy band structure of \ch{WSe_2}}
To calculate the electronic structure of the \ch{WSe_2} trilayer, we employ the Atomic Simulation Recipe (ASR) framework with GPAW as the electronic structure code.\cite{gjerding2021atomic_si}\cite{enkovaara2010electronic_si} To obtain the atomic structure, we start from the relaxed \ch{WSe_2} monolayer downloaded from the C2DB database\cite{haastrup2018computational,gjerding2021recent}, which we stack three copies of in the optimal 2H-stacking configuration. Subsequently, the structure is relaxed using the PBE-D3 scheme to account for van der Waals interactions.\cite{grimme2010consistent} 
We then preform a ground state calculation. The calculations were performed in plane-wave mode with a energy cut-off of 800 eV on a $\Gamma$-point centered $K$-point grid with a density of 12000 points/\AA$^2$. 
Spin-orbit coupling was included perturbatively. Based on the ground state calculation, we can obtain the electronic band structure, the effective masses of both electrons and holes, and the density of states, using the ASR bandstructure recipe, the ASR effective masses recipe, and the ASR pdos recipe respectively. 

We find an electronic band gap of 1.05 eV. The effective mass of the valence band maximum is isotropic at 0.36 electron masses, $m_{\rm e}$, and the effective mass of the conduction band is anisotropic with an effective mass of $0.46~m_{\rm e}$ in the fast direction and $0.58~m_{\rm e}$ in the slow direction. Based on the calculated density of states, we can estimate the effective density of states at the conduction band edge, $N_c$, and valence band edge, $N_v$,
\begin{align}
    & N_{\rm v} = \int_{-\infty}^{E_{\rm v}} dE\, \mathrm{DOS}(E)\mathrm{exp}[(E-E_{\rm v})/(kT)] = 1.3\cdot 10^{17} \,\mathrm{states \cdot m}^{-2}, \\
    & N_{\rm c} = \int_{E_{\rm c}}^\infty dE\, \mathrm{DOS}(E)\mathrm{exp}[-(E-E_{\rm c})/(kT)] = 2.2\cdot 10^{17}\, \mathrm{states \cdot m}^{-2}.
 \end{align}
To model the gating efficiency of our \ch{WSe_2} gate we approximate the density of states with two constants $g_{\rm v/c}$, below and above the gap respectively, chosen to reproduce the calculated effective densities of stats. 
These read
\begin{align}\label{eq:g_values}
g_{\rm v}=N_{\rm v}/(k{\rm B}T) =4.9\cdot 10^{18} \,\mathrm{states\cdot m^{-2} \cdot eV^{-1}}, \\
g_{\rm c} = N_{\rm c}/(k{\rm B}T)= 8.5\cdot 10^{18}\, \mathrm{states\cdot m^{-2} \cdot eV^{-1}}.
 \end{align}
\begin{figure}[h]
\centering
\includegraphics[width=0.7\textwidth]{\pathfigs 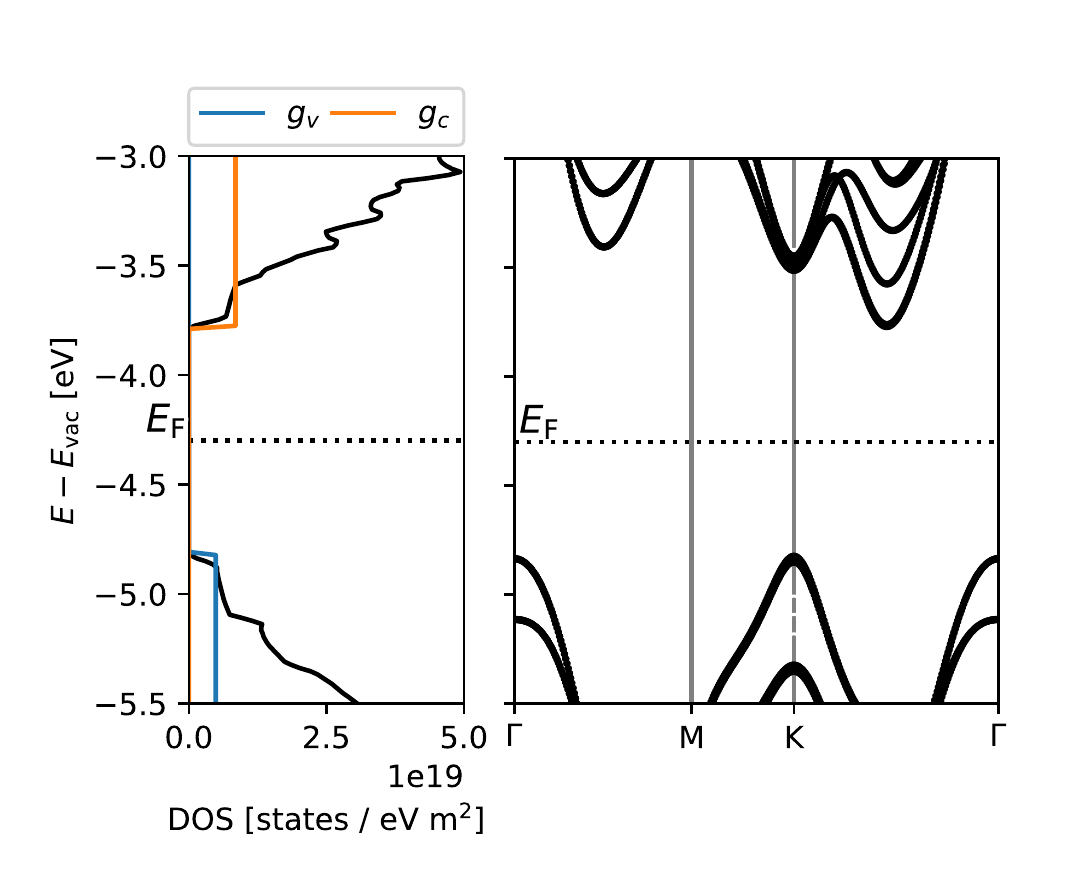}
\caption{Left: density of states for a 2H-stacked trilayer of \ch{WSe_2}. Blue and orange lines represent the approximate densities of states of the valence and conduction band respectively used in the actual modeling of the gating response. Right: DFT calculated electronic band structure for a 2H-stacked trilayer of \ch{WSe_2} in the relevant energy range along the $\Gamma MK\Gamma$ contour.}
\label{fig:bs_fig}
\end{figure}

\section{Carrier density in graphene as a function of the top and bottom gate voltages}
Here we derive a minimal model, based on electrostatics and equilibrium of {\it electrochemical} potential, to calculate the density of carriers in graphene as a function of the top and bottom gate voltages.
The main source of deviation from a purely electrostatic behavior is the quantum capacitance of the \ch{WSe_2} that becomes very small when the chemical potential is pushed deeply inside the gap due to the vanishing density of states, hence dominating over all the other capacities (that are summed in series).

In this simplified model we consider both bilayer graphene (BLG) and few-layer \ch{WSe_2} as perfectly two-dimensional materials neglecting the electrostatic potential drop between the different layers. 
To be consistent with this assumption we also neglect the modification of the BLG band structure due to the presence of an out-of-plane field and the corresponding gap-opening.

\subsection{Electrostatics}
\begin{figure}
\includegraphics[scale=1]{\pathfigs 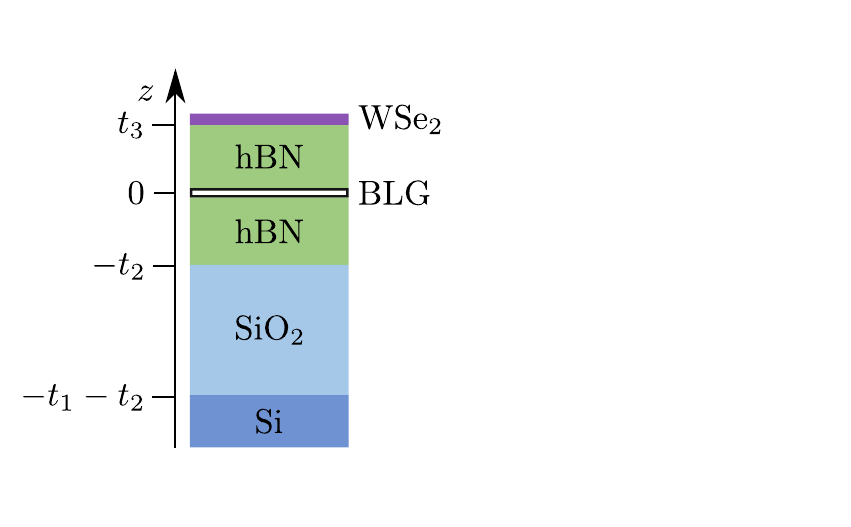}
\caption{\label{fig:scheme} Scheme of the cross-section of our device. The origin of the $z$ coordinate is taken at the position of the BLG layer. The thicknesses of the layers (out of scale) are $t_1=285~{\rm nm}$, $t_2=25~{\rm nm}$, and $t_3=4~{\rm nm}$.}
\end{figure}

We start by solving the Poisson equation in our structure. Neglecting edge effects this reads (in International System units) 
\begin{equation}\label{eq:poisson}
-\partial_z[\epsilon_0\epsilon_\perp(z) \partial_z \phi(z) ]= \rho(z),
\end{equation}
where $\phi(z)$ is the electrostatic potential as a function of the out-of plane coordinate $z$, $\rho(z)$ is the density of charges not bound in dielectrics, $\epsilon_0\approx 8.85 \cdot 10^{-12}~{\rm F/m}$ is the vacuum permittivity, and $\epsilon_\perp(z)$ is the relative dielectric permittivity of the structure.
This is given by (See Fig.~\ref{fig:scheme})
\begin{equation}\label{eq:dielectric}
\epsilon_\perp(z)=
\begin{cases}
\epsilon_{\ch{SiO_2}}\approx 4.1\quad &-t_{1}-t_{2}<z<-t_{2}\\
\epsilon_{{\rm hBN}\perp}\approx 3.5\quad& -t_{2}<z<t_{3}.
\end{cases}
\end{equation}
Since no free charge is hosted in the dielectrics we can write the charge density as
\begin{equation}\label{eq:charge_density}
\rho(z) = \rho_{\ch{WSe_2}}\delta(z-t_{3}) + \rho_{\rm BLG} \delta(z) + \rho_{\ch{Si}} \delta(z+t_{1} + t_{2}).
\end{equation}
Note that in writing (\ref{eq:charge_density}) we have neglected the charge density that can build up at the \ch{hBN}-\ch{SiO_2} interface due to photodoping.
This contribution is difficult to quantify and only results in a rigid shift of the BLG carrier density.
Since the device is overall neutral the three surface charge densities sum to zero 
\begin{equation}\label{eq:neutrality}
\rho_{\ch{WSe_2}} + \rho_{\rm BLG}  + \rho_{\ch{Si}} =0.
\end{equation}
Under the condition of vanishing electric field for $z>t_3$ and $z<-t_1-t_2$, Eq.~(\ref{eq:poisson}) can be solved yielding a continuous, piecewise linear function
\begin{equation}
\phi(z) = 
\begin{cases}
\phi_{\ch{WSe_2}}\frac{z}{t_3}+\phi_{\rm BLG}\frac{z-t_3}{-t_3}\quad 0<z<t_3\\
\phi_{\rm BLG}\frac{z+t_2}{t_2}+\phi_{\rm I}\frac{z}{-t_2}\quad -t_2<z<0\\
\phi_{\rm I} \frac{z + t_2+t_1}{t_1} + \phi_{\ch{Si}} \frac{z+t_2}{-t_1}\quad -t_1-t_2<z<-t_2,
\end{cases}
\end{equation}
where the values of the electrostatic potential at the interfaces are given by
\begin{align}
\phi_{\ch{WSe_2}} & = \frac{\rho_{\ch{WSe_2}}}{C_{\rm T}}\label{eq:wphi}\\
\phi_{\rm BLG} & = 0\label{eq:blgphi} \\
\phi_{\rm I} & = \frac{\rho_{\ch{Si}}t_2}{\epsilon_0\epsilon_{{\rm hBN}\perp}}\\
\phi_{\ch{Si}} & = \frac{\rho_{\ch{Si}}}{C_{\rm B}}.\label{eq:siphi}
\end{align}
Here, we specified the overall additive constant by requiring the electrostatic potential to vanish at the position of the BLG layer, while the two capacities are given by $C_{\rm T} = \epsilon_0\epsilon_{{\rm hBN}\perp}/t_{3}\approx 7.7 ~{\rm mF/m^2}$ and $C_{\rm B} = [(\epsilon_0\epsilon_{{\rm hBN}\perp}/t_{2})^{-1}+(\epsilon_0\epsilon_{\ch{SiO_2}}/t_{1})^{-1}]^{-1}\approx 0.12~{\rm mF/m^2}$.

\subsection{Quantum capacitances}

The equilibrium conditions between the different layers are expressed in terms of the electrochemical potential $\Phi$.
This is defined for each layer as
\begin{equation}\label{eq:electrochemical}
-e \Phi_\alpha = -e \phi_\alpha + \mu_\alpha,
\end{equation}
where $\alpha =\ch{WSe_2}, {\rm BLG}, \ch{Si}$ and $\mu_\alpha$ is the chemical potential measured from the vacuum level.
We therefore need a relation between the charge density and the chemical potential in each layer.

In silicon, due to the large density of states we can neglect the shift in chemical potential due to the induced charge density and the chemical potential stays at a fixed value given by the negative of the Silicon work function $W_{\ch{Si}}$, i.e.
\begin{equation}\label{eq:simu}
\mu_{\ch{Si}} \approx -W_{\ch{Si}}.
\end{equation}

In BLG the charge density can be expressed as 
\begin{equation}\label{eq:blgrho}
\rho_{\rm BLG} = -e(n_{\rm BLG} - \bar{n}_{\rm BLG}),
\end{equation}
where $n_{\rm BLG}$ is the density of mobile carriers that determines the plasmon dispersion and $e \bar{n}_{\rm BLG}$ is the residual charge density that is left in BLG when the Fermi level is at the Dirac point due to impurity doping.
Approximating the first valence and conduction bands of BLG as parabolic bands yields an energy independent density of states in the vicinity of the Dirac point given by $g = 2m^* /(\pi \hbar^2)$ with $m^*\approx 0.046~ m_{\rm e}$, $m_{\rm e}$ being the bare electron mass. This allows us to calculate $n_{\rm BLG}$ as
\begin{equation}\label{eq:blg_density}
n_{\rm BLG} = g\left[I(\mu_{\rm BLG}-E_{\rm D},k_{\rm B}T)-I(-\mu_{\rm BLG}+E_{\rm D},k_{\rm B}T)\right] = g(\mu_{\rm BLG}-E_{\rm D}),
\end{equation}
where $E_{\rm D}\approx -4.5 ~{\rm eV}$ is the energy of the Dirac point of BLG measured from the vacuum level, and
\begin{equation}\label{eq:I}
I(\mu, k_{\rm B} T) = \int_0^\infty \frac{dE}{1+\exp[(E-\mu)/(k_{\rm B}T)]} = k_{\rm B}T \ln\left(1+e^{\frac{\mu}{k_{\rm B}T}}\right) \stackrel{\frac{|\mu|}{k_{\rm B}T}\gg 1}{\approx}
\begin{cases}
\mu\quad\mu >0\\
k_{\rm B}T e^{\frac{\mu}{k_{\rm B}T}}\quad\mu <0.
\end{cases}
\end{equation}
Imposing that BLG is electrically neutral when the chemical potential equals the negative of the work function of isolated BLG $\bar{W}_{\rm BLG}$ (in the following bars will denote quantities related to the isolated materials) we obtain 
\begin{equation}
-eg(-\bar{W}_{\rm BLG} -E_{\rm D})+e\bar{n}_{\rm BLG}=0,
\end{equation}
that we can use to eliminate $E_{\rm D}$ in (\ref{eq:blg_density}). Solving for the chemical potential gives
\begin{equation}\label{eq:blgmu}
\mu_{\rm BLG} =-\bar{W}_{\rm BLG}+\frac{e^2}{C_{\rm Q}}(n_{\rm BLG}-\bar{n}_{\rm BLG}),
\end{equation}
where $C_{\rm Q} = e^2g\approx 62 ~{\rm mF/m^2}$ is the quantum capacitance of BLG.

Similarly, in \ch{WSe_2} we can write the charge density as
\begin{equation}\label{eq:wdensity}
\rho_{\ch{WSe_2}} = -e(n_{\ch{WSe_2}} -e\bar{n}_{\ch{WSe_2}}),
\end{equation}
where $n_{\ch{WSe_2}}$ is the number of mobile carriers in the bands (electrons-holes) and $e\bar{n}_{\ch{WSe_2}}$ is a residual charge density due to doping.
Again, approximating the relevant bands with an effective mass we get constant densities of states in the conduction and valence bands $g_{\rm C/\rm V}$ (using values from Eq. \ref{eq:g_values}), and the carrier density can be calculated according to 
\begin{equation}
n_{\ch{WSe_2}}[\mu_{\ch{WSe_2}}] = g_cI(\mu_{\ch{WSe_2}}-E_{\rm C},k_{\rm B}T)-g_vI(E_{\rm V}-\mu_{\ch{WSe_2}},k_{\rm B}T),
\end{equation}
where $E_{\rm C/V}$ are the edges of the conduction/valence band edges measured from the vacuum level.
Imposing charge neutrality when the chemical potential equals minus the work function of isolated \ch{WSe_2} gives
\begin{equation}
\bar{n}_{\ch{WSe_2}} = g_cI(-\bar{W}_{\ch{WSe_2}}-E_{\rm C},k_{\rm B}T)-g_vI(E_{\rm V}+\bar{W}_{\ch{WSe_2}},k_{\rm B}T).
\end{equation}
In the following we will express the chemical potential of \ch{WSe_2} in terms of a deviation $\Delta\bar{\mu}_{\ch{WSe_2} }$  from its value in pristine \ch{WSe_2}, i.e.
\begin{equation}\label{eq:wmu}
\mu_{\ch{WSe_2} }= -\bar{W}_{\ch{WSe_2}} +\Delta\bar{\mu}_{\ch{WSe_2} }.
\end{equation}

\subsection{Equilibrium of electrochemical potential}

The voltage source connected at the bottom gate fixes the electrochemical potential difference between BLG and silicon to be
\begin{equation}
\Phi_{\ch{Si}} - \Phi_{\rm BLG} = V_{\rm B},
\end{equation}
substituting (\ref{eq:siphi}-\ref{eq:simu}-\ref{eq:blgphi}-\ref{eq:blgmu}) we can solve for the silicon surface charge density
\begin{equation}\label{eq:sirho}
\rho_{\ch{Si}}=C_{\rm B}\left(V_{\rm B}-\bar{V}_{\rm B}-e\frac{n_{\rm BLG}-\bar{n}_{\rm BLG}}{C_{\rm Q}}\right),
\end{equation}
where $\bar{V}_{\rm B} = (W_{\ch{Si}}-\bar{W}_{\rm BLG})/e$.

In a similar way we can write for the top-gate voltage
\begin{equation}
\Phi_{\ch{WSe_2}} - \Phi_{\rm BLG}  = V_{\rm T},
\end{equation}
and, substituting (\ref{eq:wphi}-\ref{eq:wmu}-\ref{eq:blgphi}-\ref{eq:blgmu}), solve for the \ch{WSe_2} charge density
\begin{equation}\label{eq:wrho}
\rho_{\ch{WSe_2}}=C_{\rm T}\left(V_{\rm T}-\bar{V}_{\rm T}+ \frac{\Delta\bar{\mu}_{\ch{WSe_2}}}{e}-e\frac{n_{\rm BLG}-\bar{n}_{\rm BLG}}{C_{\rm Q}}\right).
\end{equation}
Here, $\bar{V}_{\rm T} = (\bar{W}_{\ch{WSe_2}}-\bar{W}_{\rm BLG})/e$.

Subsituting (\ref{eq:blgrho}-\ref{eq:sirho}-\ref{eq:wrho}) into the charge neutrality condition (\ref{eq:neutrality}) gives 
\begin{equation}
\left(1+\frac{C_{\rm T}}{C_{\rm Q}}+\frac{C_{\rm B}}{C_{\rm Q}}\right)e(n_{\rm BLG}-\bar{n}_{\rm BLG}) =C_{\rm T}(V_{\rm T}-\bar{V}_{\rm T}) + C_{\rm B} (V_{\rm B}-\bar{V}_{\rm B})+\frac{C_{\rm T}}{e}\Delta\bar{\mu}_{\ch{WSe_2}},
\end{equation}
that, together with the equation obtained by eliminating $\rho_{\ch{WSe_2}}$ from (\ref{eq:wdensity}) and (\ref{eq:wrho}),
\begin{equation}
 -en_{\ch{WSe_2}}[-\bar{W}_{\ch{WSe_2}} +\Delta\bar{\mu}_{\ch{WSe_2} }] +e\bar{n}_{\ch{WSe_2}}= C_{\rm T}\left(V_{\rm T}-\bar{V}_{\rm T}+ \frac{\Delta\bar{\mu}_{\ch{WSe_2}}}{e}-e\frac{n_{\rm BLG}-\bar{n}_{\rm BLG}}{C_{\rm Q}}\right),
\end{equation}
constitutes a system of two non-linear equations in the two variables $n_{\rm BLG}$ and $\Delta\bar{\mu}_{\ch{WSe_2}}$ that can be solved numerically.

However, before solving these equations we exploit the fact that $C_{\rm Q}\gg C_{\rm T}, C_{\rm B}$ and take the limit $C_{\rm Q}\to \infty$.
This gives 
\begin{align}
e(n_{\rm BLG}-\bar{n}_{\rm BLG}) =C_{\rm T}(V_{\rm T}-\bar{V}_{\rm T}) + C_{\rm B} (V_{\rm B}-\bar{V}_{\rm B})+\frac{C_{\rm T}}{e}\Delta\bar{\mu}_{\ch{WSe_2}},\label{eq:capacitor}\\
 -en_{\ch{WSe_2}}[-\bar{W}_{\ch{WSe_2}} +\Delta\bar{\mu}_{\ch{WSe_2} }] +e\bar{n}_{\ch{WSe_2}}= C_{\rm T}\left(V_{\rm T}-\bar{V}_{\rm T}+ \frac{\Delta\bar{\mu}_{\ch{WSe_2}}}{e}\right).\label{eq:topgate}
\end{align}
Note that in this approximation the second equation (\ref{eq:topgate}) does not depend on $V_{\rm B}$.

To simplify the analysis of experimental results it is useful to use as a reference in our equations the zero bias situation ($V_{\rm B} = V_{\rm T} = 0$) instead of the flat vacuum level situation ($V_{\rm B}=\bar {V_{\rm B}}$, $V_{\rm T} = \bar{V_{\rm T}}$). This can be done by defining $\Delta\bar{\mu}_{\ch{WSe_2}}^0$ as the solution of 
\begin{equation}\label{eq:zerobias}
 -en_{\ch{WSe_2}}[-\bar{W}_{\ch{WSe_2}} +\Delta\bar{\mu}_{\ch{WSe_2}}^0] +e\bar{n}_{\ch{WSe_2}}= C_{\rm T}\left(-\bar{V}_{\rm T}+ \frac{\Delta\bar{\mu}_{\ch{WSe_2}}^0}{e}\right).
\end{equation}
Subtracting (\ref{eq:zerobias}) from (\ref{eq:topgate}), and defining $W_{\ch{WSe_2}}^0 = \bar{W}_{\ch{WSe_2}} - \Delta\bar{\mu}_{\ch{WSe_2}}^0$, and $\Delta\mu_{\ch{WSe_2}}=\Delta \bar{\mu}_{\ch{WSe_2}}-\Delta\mu^0_{\ch{WSe_2}}$, gives our final equation for the chemical potential shift $\Delta\mu_{\ch{WSe_2}}$ 
\begin{empheq}[box=\fbox]{equation}\label{eq:chemical_potential_equation}
n_{\ch{WSe_2}}[\Delta{\mu}_{\ch{WSe_2}}-W_{\ch{WSe_2}}^0] -n_{\ch{WSe_2}}[-W_{\ch{WSe_2}}^0] = -\frac{C_{\rm T}}{e}\left(V_{\rm T}+ \frac{\Delta{\mu}_{\ch{WSe_2}}}{e}\right).
\end{empheq}
The result can be then fed into
%
\begin{empheq}[box=\fbox]{equation}\label{eq:density_equation}
n_{\rm BLG} = \frac{C_{\rm B} (V_{\rm B}-V_{\rm D})}{e}+\frac{C_{\rm T}V_{\rm T}}{e} +\frac{C_{\rm T}\Delta\mu_{\ch{WSe_2}}[V_{\rm T}]}{e^2},
\end{empheq}
that is derived from (\ref{eq:capacitor}) by defining $n_{\rm BLG}^0 = \bar{n}_{\rm BLG} - C_{\rm B} \bar{V}_{\rm B}/e-C_{\rm T}\bar{V}_{\rm T}/e +C_{\rm T}\Delta\mu_{\ch{WSe_2}}^0/e^2$, $V_{\rm D} =-en_{\rm BLG}^0/C_{\rm B}$. Note that $\Delta\mu_{\ch{WSe_2}}[V_{\rm T}=0] =0$.
\subsection{Solution of Eqs.~(\ref{eq:chemical_potential_equation}-\ref{eq:density_equation})}
To solve Eq.~(\ref{eq:chemical_potential_equation}) for $\Delta{\mu}_{\ch{WSe_2}}$ as a function of $V_{\rm T}$ requires in general the use of a numerical root finding method.
However, it is instructive to look at the zero temperature limit where an analytical solution can be derived.
Making use of (\ref{eq:I}) the left hand side of (\ref{eq:chemical_potential_equation}) simplifies, at zero temperature, to  
\begin{equation}
n_{\ch{WSe_2}}[\Delta{\mu}_{\ch{WSe_2}}-W_{\ch{WSe_2}}^0] -n_{\ch{WSe_2}}[-W_{\ch{WSe_2}}^0] =
\begin{cases}
g_{\rm V}(\Delta{\mu}_{\ch{WSe_2}}+eV_{\rm V})& \;\Delta{\mu}_{\ch{WSe_2}} <-eV_{\rm V}\\
0 &\;  -eV_{\rm V}\leq \Delta{\mu}_{\ch{WSe_2}}\leq -eV_{\rm C}\\
g_{\rm C} (\Delta{\mu}_{\ch{WSe_2}}+eV_{\rm C})&   \; \Delta{\mu}_{\ch{WSe_2}} >-eV_{\rm C},
\end{cases}
\end{equation}
where $V_{\rm C}= -(E_{\rm C}+W_{\ch{WSe_2}}^0)/e$ and $V_{\rm V} = -(E_{\rm V}+W_{\ch{WSe_2}}^0)/e$.

\begin{figure}
\includegraphics[scale=1]{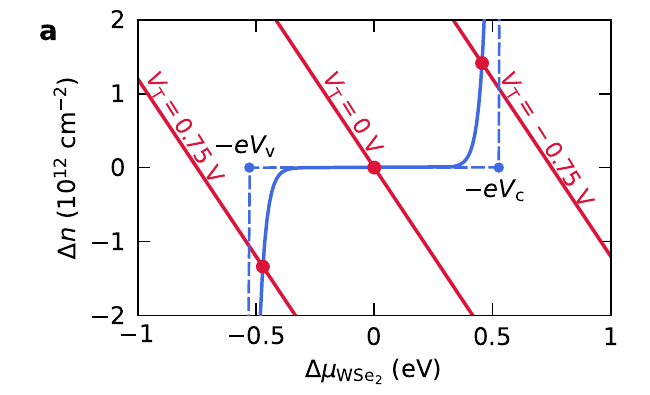}
\caption{\label{fig:solution} Graphical solution of Eq. (\ref{eq:chemical_potential_equation}). The tick blue curve represents the left hand side of (\ref{eq:chemical_potential_equation}) evaluated at room temperature $k_{\rm B} T =~ 0.026 {\rm eV}$. The thin red lines represent the the right hand side of (\ref{eq:chemical_potential_equation}) for three different values of $V_{\rm T}$. 
The dashed blue curve represents the left hand side of (\ref{eq:chemical_potential_equation}) evaluated  at zero temperature.}
\end{figure}

\begin{figure}
\includegraphics[scale=1]{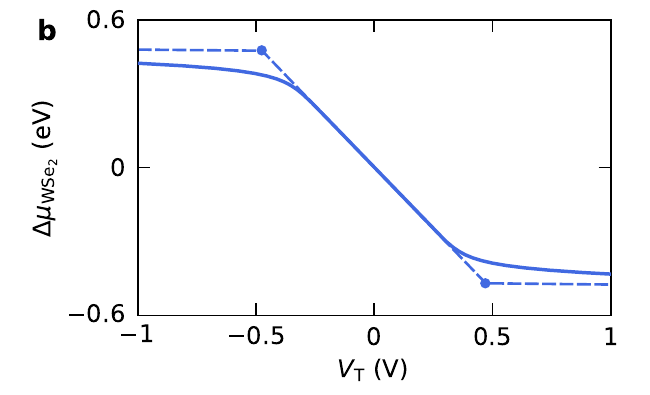}
\caption{\label{fig:delta} Solution of  Eq. (\ref{eq:chemical_potential_equation}). The thick blue line represents $\Delta\mu_{\ch{WSe_2}}$ as a function of $V_{\rm T}$ calculated by solving numerically (\ref{eq:chemical_potential_equation}). The blue dashed line represents the zero-temperature solution (\ref{eq:deltamu0t}).}
\end{figure}

\begin{figure}[h!!]
\includegraphics[scale=1]{\pathfigs 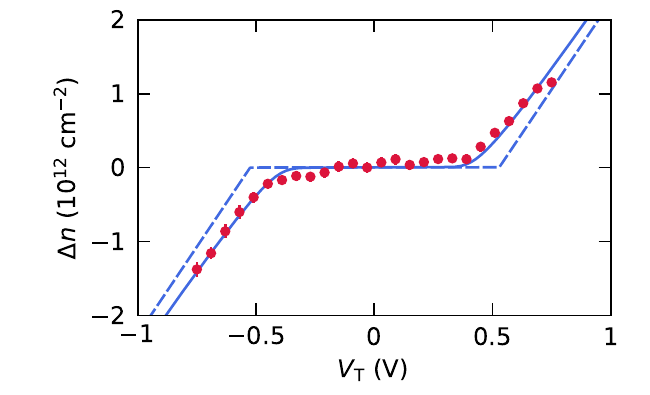}
\caption{\label{fig:compare} Carrier density induced by the application of a top-gate voltage. Dots are experimental data, the thick blue curve is the full numerical solution while the dashed blue curve corresponds to the analytical zero-temperature solution.}
\end{figure}

This expression is plotted in Figure~(\ref{fig:solution}) as dashed line together with the right-hand side of (\ref{eq:chemical_potential_equation}) calculated for three different values of $V_{\rm T}$.
The solution can be obtained by the intersection of the two curves and gives, in the zero temperature limit, a piecewise linear function
\begin{equation}\label{eq:deltamu0t}
\Delta\mu_{\ch{WSe_2}}[V_{\rm T}] = 
\begin{cases}
\frac{-eV_{\rm C}-eV_{\rm T} \frac{C_{\rm T}}{e^2g_{\rm C}} }{1+ \frac{C_{\rm T}}{e^2g_{\rm C}}}\approx -eV_{\rm C}&   \quad V_{\rm T} <V_{\rm C}\\
-eV_{\rm T} & \quad V_{\rm C}\leq V_{\rm T} \leq V_{\rm V}\\
\frac{-eV_{\rm V}-eV_{\rm T} \frac{C_{\rm T}}{e^2g_{\rm V}} }{1+ \frac{C_{\rm T}}{e^2g_{\rm V}}}\approx -eV_{\rm V}& \quad V_{\rm T} >V_{\rm V},\\
\end{cases}
\end{equation}
where we made use of $e^2 g_{{\rm C},{\rm V}}\gg C_{\rm T}$.
Figure~(\ref{fig:solution}) also shows the finite-temperature curve showing that at finite temperature solution is always closer to zero than the zero-temperature one. 
We can then compare the full numerical result with the analytical, zero-temperature approximation.
These are shown in Figure~(\ref{fig:delta}).
We see that the numerical and the analytical solutions agree qualitatively, showing a linear behavior for values of the top-gate voltage inside the \ch{WSe_2} gap and saturations at the conduction and valence band edges. However, finite temperature effects significantly reduces the value of $\Delta{\mu}_{\ch{WSe_2}}$.
From (\ref{eq:density_equation}) we can calculate the shift in density produced by the application of $V_{\rm T}$, again, in the zero temperature limit, this gives
\begin{equation}
\delta n_{\rm BLG} = n_{\rm BLG} -n_{\rm BLG}^0 - \frac{C_{\rm B} V_{\rm B}}{e} =\frac{C_{\rm T}}{e} \times \begin{cases}
V_{\rm T}-V_{\rm C} \quad & V_{\rm T}  <V_{\rm C}\\
0\quad & V_{\rm C}\leq V_{\rm T}  \leq V_{\rm V}\\
V_{\rm T} -V_{\rm V}\quad & V_{\rm T}  >V_{\rm V}.\\
\end{cases}
\end{equation}
And we compare the analytical solution with the full numerical solution and experimental data in Figure (\ref{fig:compare}).


\begin{thebibliography}{10}
\expandafter\ifx\csname url\endcsname\relax
  \def\url#1{\texttt{#1}}\fi
\expandafter\ifx\csname urlprefix\endcsname\relax\def\urlprefix{}\fi
\providecommand{\bibinfo}[2]{#2}
\providecommand{\eprint}[2][]{\url{#2}}

\bibitem{novotny2006near}
\bibinfo{author}{Novotny, L.} \& \bibinfo{author}{Stranick, S.~J.}
\newblock Near-field optical microscopy and spectroscopy with pointed probes.
\newblock \textit{\bibinfo{journal}{Annu. Rev. Phys. Chem.}}
  \textbf{\bibinfo{volume}{57}}, \bibinfo{pages}{303--331}
  (\bibinfo{year}{2006}).

\bibitem{Chen2012}
\bibinfo{author}{Chen, J.} et~al.
\newblock \href{http://dx.doi.org/10.1038/nature11254}{{Optical nano-imaging of
  gate-tunable graphene plasmons}}.
\newblock \textit{\bibinfo{journal}{Nature}} \textbf{\bibinfo{volume}{487}},
  \bibinfo{pages}{77--81} (\bibinfo{year}{2012}).

\bibitem{Fei2012}
\bibinfo{author}{Fei, Z.} et~al.
\newblock \href{http://dx.doi.org/10.1038/nature11253}{{Gate-tuning of graphene
  plasmons revealed by infrared nano-imaging}}.
\newblock \textit{\bibinfo{journal}{Nature}} \textbf{\bibinfo{volume}{487}},
  \bibinfo{pages}{82--85} (\bibinfo{year}{2012}).

\bibitem{Dai2014}
\bibinfo{author}{Dai, S.} et~al.
\newblock \href{http://dx.doi.org/10.1126/science.1246833}{{Tunable Phonon
  Polaritons in Atomically Thin van der Waals Crystals of Boron Nitride}}.
\newblock \textit{\bibinfo{journal}{Science}} \textbf{\bibinfo{volume}{343}},
  \bibinfo{pages}{1125--1129} (\bibinfo{year}{2014}).

\bibitem{BasovNanophotonics2021}
\bibinfo{author}{Basov, D.~N.}, \bibinfo{author}{Asenjo-Garcia, A.},
  \bibinfo{author}{Schuck, P.~J.}, \bibinfo{author}{Zhu, X.} \&
  \bibinfo{author}{Rubio, A.}
\newblock \href{http://dx.doi.org/doi:10.1515/nanoph-2020-0449}{Polariton
  panorama}.
\newblock \textit{\bibinfo{journal}{Nanophotonics}}
  \textbf{\bibinfo{volume}{10}}, \bibinfo{pages}{549--577}
  (\bibinfo{year}{2021}).

\bibitem{Basov2016}
\bibinfo{author}{Basov, D.~N.}, \bibinfo{author}{Fogler, M.~M.} \&
  \bibinfo{author}{{Garcia de Abajo}, F.~J.}
\newblock \href{http://dx.doi.org/10.1126/science.aag1992}{{Polaritons in van
  der Waals materials}}.
\newblock \textit{\bibinfo{journal}{Science}} \textbf{\bibinfo{volume}{354}}
  (\bibinfo{year}{2016}).

\bibitem{Low2017}
\bibinfo{author}{Low, T.} et~al.
\newblock \href{http://dx.doi.org/10.1038/nmat4792}{{Polaritons in layered
  two-dimensional materials}}.
\newblock \textit{\bibinfo{journal}{Nature Materials}}
  \textbf{\bibinfo{volume}{16}}, \bibinfo{pages}{182--194}
  (\bibinfo{year}{2017}).

\bibitem{Woessner2016}
\bibinfo{author}{Woessner, A.} et~al.
\newblock \href{http://dx.doi.org/10.1038/ncomms10783}{{Near-field photocurrent
  nanoscopy on bare and encapsulated graphene}}.
\newblock \textit{\bibinfo{journal}{Nature Communications}}
  \textbf{\bibinfo{volume}{7}}, \bibinfo{pages}{10783} (\bibinfo{year}{2016}).

\bibitem{Woessner2015}
\bibinfo{author}{Woessner, A.} et~al.
\newblock \href{http://dx.doi.org/10.1038/nmat4169}{{Highly confined low-loss
  plasmons in graphene–boron nitride heterostructures}}.
\newblock \textit{\bibinfo{journal}{Nature Materials}}
  \textbf{\bibinfo{volume}{14}}, \bibinfo{pages}{421--425}
  (\bibinfo{year}{2015}).

\bibitem{Ni2018}
\bibinfo{author}{Ni, G.~X.} et~al.
\newblock \href{http://dx.doi.org/10.1038/s41586-018-0136-9}{{Fundamental
  limits to graphene plasmonics}}.
\newblock \textit{\bibinfo{journal}{Nature}} \textbf{\bibinfo{volume}{557}},
  \bibinfo{pages}{530--533} (\bibinfo{year}{2018}).

\bibitem{Lundeberg2017}
\bibinfo{author}{Lundeberg, M.~B.} et~al.
\newblock \href{http://dx.doi.org/10.1126/science.aan2735}{{Tuning quantum
  nonlocal effects in graphene plasmonics}}.
\newblock \textit{\bibinfo{journal}{Science}} \textbf{\bibinfo{volume}{357}},
  \bibinfo{pages}{187--191} (\bibinfo{year}{2017}).

\bibitem{Sunku2018}
\bibinfo{author}{Sunku, S.~S.} et~al.
\newblock \href{http://dx.doi.org/10.1126/science.aau5144}{{Photonic crystals
  for nano-light in moir{\'{e}} graphene superlattices}}.
\newblock \textit{\bibinfo{journal}{Science}} \textbf{\bibinfo{volume}{362}},
  \bibinfo{pages}{1153--1156} (\bibinfo{year}{2018}).

\bibitem{Schmidt2018}
\bibinfo{author}{Schmidt, P.} et~al.
\newblock \href{http://dx.doi.org/10.1038/s41565-018-0233-9}{{Nano-imaging of
  intersubband transitions in van der Waals quantum wells}}.
\newblock \textit{\bibinfo{journal}{Nature Nanotechnology}}
  \textbf{\bibinfo{volume}{13}}, \bibinfo{pages}{1035--1041}
  (\bibinfo{year}{2018}).

\bibitem{McCann2013}
\bibinfo{author}{McCann, E.} \& \bibinfo{author}{Koshino, M.}
\newblock \href{http://dx.doi.org/10.1088/0034-4885/76/5/056503}{{The
  electronic properties of bilayer graphene}}.
\newblock \textit{\bibinfo{journal}{Reports on Progress in Physics}}
  \textbf{\bibinfo{volume}{76}}, \bibinfo{pages}{056503}
  (\bibinfo{year}{2013}).

\bibitem{Hasdeo2017}
\bibinfo{author}{Hasdeo, E.~H.} \& \bibinfo{author}{Song, J. C.~W.}
\newblock \href{http://dx.doi.org/10.1021/acs.nanolett.7b02584}{{Long-Lived
  Domain Wall Plasmons in Gapped Bilayer Graphene}}.
\newblock \textit{\bibinfo{journal}{Nano Letters}}
  \textbf{\bibinfo{volume}{17}}, \bibinfo{pages}{7252--7257}
  (\bibinfo{year}{2017}).

\bibitem{Liu2020}
\bibinfo{author}{Liu, X.} et~al.
\newblock \href{http://dx.doi.org/10.1038/s41586-020-2458-7}{{Tunable
  spin-polarized correlated states in twisted double bilayer graphene}}.
\newblock \textit{\bibinfo{journal}{Nature}} \textbf{\bibinfo{volume}{583}},
  \bibinfo{pages}{221--225} (\bibinfo{year}{2020}).

\bibitem{Cao2020}
\bibinfo{author}{Cao, Y.} et~al.
\newblock \href{http://dx.doi.org/10.1038/s41586-020-2260-6}{{Tunable
  correlated states and spin-polarized phases in twisted bilayer–bilayer
  graphene}}.
\newblock \textit{\bibinfo{journal}{Nature}} \textbf{\bibinfo{volume}{583}},
  \bibinfo{pages}{215--220} (\bibinfo{year}{2020}).

\bibitem{Park2021}
\bibinfo{author}{Park, J.~M.}, \bibinfo{author}{Cao, Y.},
  \bibinfo{author}{Watanabe, K.}, \bibinfo{author}{Taniguchi, T.} \&
  \bibinfo{author}{Jarillo-Herrero, P.}
\newblock \href{http://dx.doi.org/10.1038/s41586-021-03192-0}{{Tunable strongly
  coupled superconductivity in magic-angle twisted trilayer graphene}}.
\newblock \textit{\bibinfo{journal}{Nature}} \textbf{\bibinfo{volume}{590}},
  \bibinfo{pages}{249--255} (\bibinfo{year}{2021}).

\bibitem{Hao2021}
\bibinfo{author}{Hao, Z.} et~al.
\newblock \href{http://dx.doi.org/10.1126/science.abg0399}{{Electric
  field–tunable superconductivity in alternating-twist magic-angle trilayer
  graphene}}.
\newblock \textit{\bibinfo{journal}{Science}} \textbf{\bibinfo{volume}{371}},
  \bibinfo{pages}{1133--1138} (\bibinfo{year}{2021}).

\bibitem{Rahman2016}
\bibinfo{author}{Rahman, F.}
\newblock \href{http://dx.doi.org/10.1201/b11137}{{Nanostructures in
  Electronics and Photonics}} (\bibinfo{publisher}{Pan Stanford},
  \bibinfo{year}{2016}).

\bibitem{AMALRICPOPESCU2001139}
\bibinfo{author}{Amalric-Popescu, D.} \& \bibinfo{author}{Bozon-Verduraz, F.}
\newblock
  \href{http://dx.doi.org/https://doi.org/10.1016/S0920-5861(01)00414-X}{Infrared
  studies on sno2 and pd/sno2}.
\newblock \textit{\bibinfo{journal}{Catalysis Today}}
  \textbf{\bibinfo{volume}{70}}, \bibinfo{pages}{139--154}
  (\bibinfo{year}{2001}).
\newblock \bibinfo{note}{A tribute to Jean-Claude Lavalley}.

\bibitem{Li2020}
\bibinfo{author}{Li, H.} et~al.
\newblock \href{http://dx.doi.org/10.1021/acs.nanolett.9b05092}{{Global Control
  of Stacking-Order Phase Transition by Doping and Electric Field in Few-Layer
  Graphene}}.
\newblock \textit{\bibinfo{journal}{Nano Letters}}
  \textbf{\bibinfo{volume}{20}}, \bibinfo{pages}{3106--3112}
  (\bibinfo{year}{2020}).

\bibitem{Sunku2021}
\bibinfo{author}{Sunku, S.~S.} et~al.
\newblock \href{http://dx.doi.org/10.1021/acs.nanolett.0c04494}{{Dual-Gated
  Graphene Devices for Near-Field Nano-imaging}}.
\newblock \textit{\bibinfo{journal}{Nano Letters}}
  \textbf{\bibinfo{volume}{21}}, \bibinfo{pages}{1688--1693}
  (\bibinfo{year}{2021}).

\bibitem{Luo2021}
\bibinfo{author}{Luo, W.} et~al.
\newblock \href{http://dx.doi.org/10.1021/acs.nanolett.1c01167}{{Nanoinfrared
  Characterization of Bilayer Graphene Conductivity under Dual-Gate Tuning}}.
\newblock \textit{\bibinfo{journal}{Nano Letters}}
  \textbf{\bibinfo{volume}{21}}, \bibinfo{pages}{5151--5157}
  (\bibinfo{year}{2021}).

\bibitem{Kim2017_EF_pinning}
\bibinfo{author}{Kim, C.} et~al.
\newblock \href{http://dx.doi.org/10.1021/acsnano.6b07159}{{Fermi Level Pinning
  at Electrical Metal Contacts of Monolayer Molybdenum Dichalcogenides}}.
\newblock \textit{\bibinfo{journal}{ACS Nano}} \textbf{\bibinfo{volume}{11}},
  \bibinfo{pages}{1588--1596} (\bibinfo{year}{2017}).

\bibitem{Movva2018}
\bibinfo{author}{Movva, H.~C.} et~al.
\newblock \href{http://dx.doi.org/10.1103/PhysRevLett.120.107703}{{Tunable
  $\Gamma$-$K$ Valley Populations in Hole-Doped Trilayer WSe$_2$}}.
\newblock \textit{\bibinfo{journal}{Physical Review Letters}}
  \textbf{\bibinfo{volume}{120}}, \bibinfo{pages}{107703}
  (\bibinfo{year}{2018}).

\bibitem{Movva2015}
\bibinfo{author}{Movva, H. C.~P.} et~al.
\newblock \href{http://dx.doi.org/10.1021/acsnano.5b04611}{{High-Mobility Holes
  in Dual-Gated WSe$_2$ Field-Effect Transistors}}.
\newblock \textit{\bibinfo{journal}{ACS Nano}} \textbf{\bibinfo{volume}{9}},
  \bibinfo{pages}{10402--10410} (\bibinfo{year}{2015}).

\bibitem{Johnson1972}
\bibinfo{author}{Johnson, P.~B.} \& \bibinfo{author}{Christy, R.~W.}
\newblock \href{http://dx.doi.org/10.1103/PhysRevB.6.4370}{Optical constants of
  the noble metals}.
\newblock \textit{\bibinfo{journal}{Phys. Rev. B}}
  \textbf{\bibinfo{volume}{6}}, \bibinfo{pages}{4370--4379}
  (\bibinfo{year}{1972}).

\bibitem{GoncalvesPeres_Book}
\bibinfo{author}{{Gon\c{c}alves}, P. A.~D.} \& \bibinfo{author}{{Peres}, N.
  M.~R.}
\newblock An introduction to graphene plasmonics (\bibinfo{publisher}{World
  Scientific}, \bibinfo{address}{Singapore}, \bibinfo{year}{2016}).

\bibitem{Podzorov2004}
\bibinfo{author}{Podzorov, V.}, \bibinfo{author}{Gershenson, M.~E.},
  \bibinfo{author}{Kloc, C.}, \bibinfo{author}{Zeis, R.} \&
  \bibinfo{author}{Bucher, E.}
\newblock \href{http://dx.doi.org/10.1063/1.1723695}{{High-mobility
  field-effect transistors based on transition metal dichalcogenides}}.
\newblock \textit{\bibinfo{journal}{Applied Physics Letters}}
  \textbf{\bibinfo{volume}{84}}, \bibinfo{pages}{3301--3303}
  (\bibinfo{year}{2004}).

\bibitem{Fang2012}
\bibinfo{author}{Fang, H.} et~al.
\newblock \href{http://dx.doi.org/10.1021/nl301702r}{{High-Performance Single
  Layered WSe$_2$ p-FETs with Chemically Doped Contacts}}.
\newblock \textit{\bibinfo{journal}{Nano Letters}}
  \textbf{\bibinfo{volume}{12}}, \bibinfo{pages}{3788--3792}
  (\bibinfo{year}{2012}).

\bibitem{Wang2018}
\bibinfo{author}{Wang, Z.} et~al.
\newblock \href{http://dx.doi.org/10.1038/s41427-018-0062-1}{{The ambipolar
  transport behavior of WSe$_2$ transistors and its analogue circuits}}.
\newblock \textit{\bibinfo{journal}{NPG Asia Materials}}
  \textbf{\bibinfo{volume}{10}}, \bibinfo{pages}{703--712}
  (\bibinfo{year}{2018}).

\bibitem{Das2013}
\bibinfo{author}{Das, S.} \& \bibinfo{author}{Appenzeller, J.}
\newblock \href{http://dx.doi.org/10.1063/1.4820408}{{WSe$_2$ field effect
  transistors with enhanced ambipolar characteristics}}.
\newblock \textit{\bibinfo{journal}{Applied Physics Letters}}
  \textbf{\bibinfo{volume}{103}}, \bibinfo{pages}{103501}
  (\bibinfo{year}{2013}).

\bibitem{Chuang2016}
\bibinfo{author}{Chuang, H.-J.} et~al.
\newblock \href{http://dx.doi.org/10.1021/acs.nanolett.5b05066}{Low-resistance
  2d/2d ohmic contacts: A universal approach to high-performance wse2, mos2,
  and mose2 transistors}.
\newblock \textit{\bibinfo{journal}{Nano Letters}}
  \textbf{\bibinfo{volume}{16}}, \bibinfo{pages}{1896--1902}
  (\bibinfo{year}{2016}).
\newblock \bibinfo{note}{PMID: 26844954}.

\bibitem{gammelgaard2021}
\bibinfo{author}{Gammelgaard, L.}, \bibinfo{author}{Whelan, P.~R.},
  \bibinfo{author}{Booth, T.~J.} \& \bibinfo{author}{Bøggild, P.}
\newblock \href{http://dx.doi.org/10.1039/D1NR05413A}{Long-term stability and
  tree-ring oxidation of wse2 using phase-contrast afm}.
\newblock \textit{\bibinfo{journal}{Nanoscale}} \textbf{\bibinfo{volume}{13}},
  \bibinfo{pages}{19238--19246} (\bibinfo{year}{2021}).

\bibitem{Ju2014}
\bibinfo{author}{Ju, L.} et~al.
\newblock \href{http://dx.doi.org/10.1038/nnano.2014.60}{{Photoinduced doping
  in heterostructures of graphene and boron nitride}}.
\newblock \textit{\bibinfo{journal}{Nature Nanotechnology}}
  \textbf{\bibinfo{volume}{9}}, \bibinfo{pages}{348--352}
  (\bibinfo{year}{2014}).

\bibitem{Low2014}
\bibinfo{author}{Low, T.}, \bibinfo{author}{Guinea, F.}, \bibinfo{author}{Yan,
  H.}, \bibinfo{author}{Xia, F.} \& \bibinfo{author}{Avouris, P.}
\newblock \href{http://dx.doi.org/10.1103/PhysRevLett.112.116801}{{Novel
  Midinfrared Plasmonic Properties of Bilayer Graphene}}.
\newblock \textit{\bibinfo{journal}{Physical Review Letters}}
  \textbf{\bibinfo{volume}{112}}, \bibinfo{pages}{116801}
  (\bibinfo{year}{2014}).

\bibitem{gjerding2021atomic}
\bibinfo{author}{Gjerding, M.} et~al.
\newblock Atomic simulation recipes: A python framework and library for
  automated workflows.
\newblock \textit{\bibinfo{journal}{Computational Materials Science}}
  \textbf{\bibinfo{volume}{199}}, \bibinfo{pages}{110731}
  (\bibinfo{year}{2021}).

\bibitem{enkovaara2010electronic}
\bibinfo{author}{Enkovaara, J.} et~al.
\newblock Electronic structure calculations with gpaw: a real-space
  implementation of the projector augmented-wave method.
\newblock \textit{\bibinfo{journal}{Journal of physics: Condensed matter}}
  \textbf{\bibinfo{volume}{22}}, \bibinfo{pages}{253202}
  (\bibinfo{year}{2010}).

\bibitem{Dai2015_bandstr}
\bibinfo{author}{Dai, X.}, \bibinfo{author}{Li, W.}, \bibinfo{author}{Wang,
  T.}, \bibinfo{author}{Wang, X.} \& \bibinfo{author}{Zhai, C.}
\newblock \href{http://dx.doi.org/10.1063/1.4907315}{{Bandstructure modulation
  of two-dimensional WSe$_2$ by electric field}}.
\newblock \textit{\bibinfo{journal}{Journal of Applied Physics}}
  \textbf{\bibinfo{volume}{117}} (\bibinfo{year}{2015}).

\bibitem{Javaid2018}
\bibinfo{author}{Javaid, M.}, \bibinfo{author}{Russo, S.~P.},
  \bibinfo{author}{Kalantar-Zadeh, K.}, \bibinfo{author}{Greentree, A.~D.} \&
  \bibinfo{author}{Drumm, D.~W.}
\newblock \href{http://dx.doi.org/10.1088/2516-1075/aadf44}{{Band structure and
  giant Stark effect in two-dimensional transition-metal dichalcogenides}}.
\newblock \textit{\bibinfo{journal}{Electronic Structure}}
  \textbf{\bibinfo{volume}{1}}, \bibinfo{pages}{015005} (\bibinfo{year}{2018}).

\bibitem{Zhao2013}
\bibinfo{author}{Zhao, W.} et~al.
\newblock \href{http://dx.doi.org/10.1021/nn305275h}{{Evolution of Electronic
  Structure in Atomically Thin Sheets of WS$_2$ and WSe$_2$}}.
\newblock \textit{\bibinfo{journal}{ACS Nano}} \textbf{\bibinfo{volume}{7}},
  \bibinfo{pages}{791--797} (\bibinfo{year}{2013}).

\bibitem{Zeng2013}
\bibinfo{author}{Zeng, H.} et~al.
\newblock \href{http://dx.doi.org/10.1038/srep01608}{{Optical signature of
  symmetry variations and spin-valley coupling in atomically thin tungsten
  dichalcogenides}}.
\newblock \textit{\bibinfo{journal}{Scientific Reports}}
  \textbf{\bibinfo{volume}{3}}, \bibinfo{pages}{1608} (\bibinfo{year}{2013}).

\bibitem{Lin2014}
\bibinfo{author}{Lin, Y.-C.} et~al.
\newblock \href{http://dx.doi.org/10.1021/nn5003858}{{Direct Synthesis of van
  der Waals Solids}}.
\newblock \textit{\bibinfo{journal}{ACS Nano}} \textbf{\bibinfo{volume}{8}},
  \bibinfo{pages}{3715--3723} (\bibinfo{year}{2014}).

\bibitem{Wang2014}
\bibinfo{author}{Wang, X.} et~al.
\newblock \href{http://dx.doi.org/10.1021/nn501175k}{{Chemical Vapor Deposition
  Growth of Crystalline Monolayer MoSe$_2$}}.
\newblock \textit{\bibinfo{journal}{ACS Nano}} \textbf{\bibinfo{volume}{8}},
  \bibinfo{pages}{5125--5131} (\bibinfo{year}{2014}).

\bibitem{Eichfeld2015}
\bibinfo{author}{Eichfeld, S.~M.} et~al.
\newblock \href{http://dx.doi.org/10.1021/nn5073286}{{Highly Scalable,
  Atomically Thin WSe$_2$ Grown via Metal–Organic Chemical Vapor
  Deposition}}.
\newblock \textit{\bibinfo{journal}{ACS Nano}} \textbf{\bibinfo{volume}{9}},
  \bibinfo{pages}{2080--2087} (\bibinfo{year}{2015}).

\bibitem{Liu2015}
\bibinfo{author}{Liu, B.} et~al.
\newblock \href{http://dx.doi.org/10.1021/acsnano.5b01301}{{Chemical Vapor
  Deposition Growth of Monolayer WSe$_2$ with Tunable Device Characteristics
  and Growth Mechanism Study}}.
\newblock \textit{\bibinfo{journal}{ACS Nano}} \textbf{\bibinfo{volume}{9}},
  \bibinfo{pages}{6119--6127} (\bibinfo{year}{2015}).

\bibitem{morpurgo2021}
\bibinfo{author}{Gutiérrez-Lezama, I.}, \bibinfo{author}{Ubrig, N.},
  \bibinfo{author}{Ponomarev, E.} \& \bibinfo{author}{Morpurgo, A.~F.}
\newblock \href{http://dx.doi.org/10.1038/s42254-021-00317-2}{Ionic gate
  spectroscopy of 2d semiconductors}.
\newblock \textit{\bibinfo{journal}{Nature Reviews Physics}}
  \textbf{\bibinfo{volume}{3}}, \bibinfo{pages}{508} (\bibinfo{year}{2021}).

\bibitem{Zomer2014}
\bibinfo{author}{Zomer, P.~J.}, \bibinfo{author}{Guimar{\~{a}}es, M. H.~D.},
  \bibinfo{author}{Brant, J.~C.}, \bibinfo{author}{Tombros, N.} \&
  \bibinfo{author}{van Wees, B.~J.}
\newblock \href{http://dx.doi.org/10.1063/1.4886096}{{Fast pick up technique
  for high quality heterostructures of bilayer graphene and hexagonal boron
  nitride}}.
\newblock \textit{\bibinfo{journal}{Applied Physics Letters}}
  \textbf{\bibinfo{volume}{105}}, \bibinfo{pages}{013101}
  (\bibinfo{year}{2014}).

\bibitem{Purdie2018}
\bibinfo{author}{Purdie, D.~G.} et~al.
\newblock \href{http://dx.doi.org/10.1038/s41467-018-07558-3}{{Cleaning
  interfaces in layered materials heterostructures}}.
\newblock \textit{\bibinfo{journal}{Nature Communications}}
  \textbf{\bibinfo{volume}{9}}, \bibinfo{pages}{5387} (\bibinfo{year}{2018}).

\bibitem{Wang2013}
\bibinfo{author}{Wang, L.} et~al.
\newblock \href{http://dx.doi.org/10.1126/science.1244358}{{One-Dimensional
  Electrical Contact to a Two-Dimensional Material}}.
\newblock \textit{\bibinfo{journal}{Science}} \textbf{\bibinfo{volume}{342}},
  \bibinfo{pages}{614--617} (\bibinfo{year}{2013}).

\bibitem{Goossens2012}
\bibinfo{author}{Goossens, A.~M.} et~al.
\newblock \href{http://dx.doi.org/10.1063/1.3685504}{{Mechanical cleaning of
  graphene}}.
\newblock \textit{\bibinfo{journal}{Applied Physics Letters}}
  \textbf{\bibinfo{volume}{100}}, \bibinfo{pages}{073110}
  (\bibinfo{year}{2012}).

\bibitem{Ocelic2006}
\bibinfo{author}{Ocelic, N.}, \bibinfo{author}{Huber, A.} \&
  \bibinfo{author}{Hillenbrand, R.}
\newblock \href{http://dx.doi.org/10.1063/1.2348781}{{Pseudoheterodyne
  detection for background-free near-field spectroscopy}}.
\newblock \textit{\bibinfo{journal}{Applied Physics Letters}}
  \textbf{\bibinfo{volume}{89}}, \bibinfo{pages}{101124}
  (\bibinfo{year}{2006}).

\end{thebibliography}

\begin{thebibliography}{1}
\bibitem{gjerding2021atomic_si}
\bibinfo{author}{Gjerding, M.} et~al.
\newblock Atomic simulation recipes: A python framework and library for
  automated workflows.
\newblock \textit{\bibinfo{journal}{Computational Materials Science}}
  \textbf{\bibinfo{volume}{199}}, \bibinfo{pages}{110731}
  (\bibinfo{year}{2021}).

\bibitem{enkovaara2010electronic_si}
\bibinfo{author}{Enkovaara, J.} et~al.
\newblock Electronic structure calculations with gpaw: a real-space
  implementation of the projector augmented-wave method.
\newblock \textit{\bibinfo{journal}{Journal of physics: Condensed matter}}
  \textbf{\bibinfo{volume}{22}}, \bibinfo{pages}{253202}
  (\bibinfo{year}{2010}).
  
\bibitem{haastrup2018computational}
\bibinfo{author}{Haastrup, S.} et~al.
\newblock The computational 2d materials database: high-throughput modeling and
  discovery of atomically thin crystals.
\newblock \textit{\bibinfo{journal}{2D Materials}}
  \textbf{\bibinfo{volume}{5}}, \bibinfo{pages}{042002} (\bibinfo{year}{2018}).

\bibitem{gjerding2021recent}
\bibinfo{author}{Gjerding, M.~N.} et~al.
\newblock Recent progress of the computational 2d materials database (c2db).
\newblock \textit{\bibinfo{journal}{2D Materials}}
  \textbf{\bibinfo{volume}{8}}, \bibinfo{pages}{044002} (\bibinfo{year}{2021}).

\bibitem{grimme2010consistent}
\bibinfo{author}{Grimme, S.}, \bibinfo{author}{Antony, J.},
  \bibinfo{author}{Ehrlich, S.} \& \bibinfo{author}{Krieg, H.}
\newblock A consistent and accurate ab initio parametrization of density
  functional dispersion correction (dft-d) for the 94 elements h-pu.
\newblock \textit{\bibinfo{journal}{The Journal of chemical physics}}
  \textbf{\bibinfo{volume}{132}}, \bibinfo{pages}{154104}
  (\bibinfo{year}{2010}).

\end{thebibliography}

\end{document}